\documentclass[12pt, authoryear]{elsarticle}

\usepackage{amssymb}
\usepackage{amsthm}
\usepackage{amsmath}
\usepackage{amsfonts}
\usepackage{mathtools}
\usepackage[inline]{enumitem}
\usepackage[hang,flushmargin]{footmisc}
\usepackage{theoremref}
\usepackage{xcolor}
\usepackage{setspace}
\usepackage{stmaryrd}

\usepackage{booktabs}
\usepackage[nolists]{endfloat}
\usepackage{threeparttable}
\usepackage{caption}
\usepackage{graphicx}

\DeclareDelayedFloatFlavor{landscape}{table}

\makeatletter
\newcommand{\inlineitem}[1][]{%
\ifnum\enit@type=\tw@
    {\descriptionlabel{#1}}
  \hspace{\labelsep}%
\else
  \ifnum\enit@type=\z@
       \refstepcounter{\@listctr}\fi
    \quad\@itemlabel\hspace{\labelsep}%
\fi}
\makeatother

\usepackage[left=2.5cm, right=2.5cm, top=3cm]{geometry}
\allowdisplaybreaks

\usepackage{titlesec}
\titleformat*{\section}{\Large\bfseries}
\titleformat*{\subsection}{\large\bfseries}
\titleformat*{\subsubsection}{\bfseries}
\titlespacing*{\section}{0pt}{0.3\baselineskip}{0.3\baselineskip}
\titlespacing*{\subsection}{0pt}{0.3\baselineskip}{0.3\baselineskip}
\titlespacing*{\subsubsection}{0pt}{0.3\baselineskip}{0.3\baselineskip}
\newtheorem{theorem}{Theorem}
\newtheorem{proposition}{Proposition}
\newtheorem{lemma}{Lemma}
\newtheorem{corollary}{Corollary}[]

\DeclareMathOperator*{\argmax}{arg\,max}

\DeclareMathOperator{\Tr}{Tr}

\makeatletter
\newcommand{\mylabel}[2]{#2\def\@currentlabel{#2}\label{#1}}
\makeatother

\renewcommand*{\thefootnote}{\fnsymbol{footnote}}

\begin{document}

\thispagestyle{empty}

\begin{center}
{\Large Estimating high-dimensional Markov-switching VARs\footnote[1]{I am grateful to Bin Chen and Nese Yildiz for their invaluable advice and encouragement. I thank Neil Ericsson, David Hendry, and seminar participants at the 23rd Dynamic Econometrics conference and the 2021 China Meeting of the Econometric Society for their useful comments and discussion. Any remaining errors are solely mine.}}\bigskip

\bigskip

$%
\begin{array}{c}
\text{{\large Kenwin Maung}} \\ 
\text{University of Rochester} \\ 
\end{array}%
$\bigskip

\today

\bigskip
\end{center}

\textit{Abstract: } {\small Maximum likelihood estimation of large Markov-switching vector autoregressions (MS-VARs) can be challenging or infeasible due to parameter proliferation. To accommodate situations where dimensionality may be of comparable order to or exceeds the sample size, we adopt a sparse framework and propose two penalized maximum likelihood estimators with either the Lasso or the smoothly clipped absolute deviation (SCAD) penalty. We show that both estimators are estimation consistent, while the SCAD estimator also selects relevant parameters with probability approaching one. A modified EM-algorithm is developed for the case of Gaussian errors and simulations show that the algorithm exhibits desirable finite sample performance. In an application to short-horizon return predictability in the US, we estimate a 15 variable 2-state MS-VAR(1) and obtain the often reported counter-cyclicality in predictability. The variable selection property of our estimators helps to identify predictors that contribute strongly to predictability during economic contractions but are otherwise irrelevant in expansions. Furthermore, out-of-sample analyses indicate that large MS-VARs can significantly outperform "hard-to-beat" predictors like the historical average.       
}

\bigskip


\smallskip


\renewcommand*{\thefootnote}{\arabic{footnote}}
\setcounter{footnote}{0}
\pagebreak

\section{Model}

We consider the asymptotic properties of the maximum likelihood estimator (MLE) for high-dimensional Markov-switching (MS) vector autoregressive (VAR) models in a double asymptotic framework where both the sample size and number of parameters are allowed to diverge. Specifically, we study the following:
\begin{equation}\label{mainmodel}
y_t = \sum_{j=1}^{p_T}A_{j,T}(S_t)y_{t-j} + \sum_{j=1}^{q_T} B_{j,T}(S_t)x_{t-j}+ \varepsilon_t ,
\end{equation}
where $y_t \in \mathbb{R}^{d_T}$ is a vector of endogenous time series, $x_t \in \mathbb{R}^{d^*_T}$ allows for the inclusion of exogenous variables, and $\varepsilon_t$ is independently and identically distributed with mean zero and positive definite variance-covariance matrix $\Sigma_T(S_t)$. The parameters $A_{j,T}(S_t)$, $B_{j,T}(S_t)$, and $\Sigma_T(S_t)$ are respectively $d_T \times d_T$, $d_T \times d_T^*$, and $d_T \times d_T$ matrices that depend on an underlying state $\{S_t\}$, which is modeled as a latent first-order Markov chain on a finite and discrete state space taking values from $\{1 , \ldots, M\}$ with $M$ known and fixed. The transition probabilities of the chain are given by $P(S_t = j| S_{t-1} = i) = p_{i\shortrightarrow j}$, and initial distributions are given by $P(S_1 = j) = p_{j}$. We define the precision matrix to be the inverse of the variance-covariance matrix, $Q_T(S_t) \equiv \Sigma_T(S_t)^{-1}$.

By including the transition matrix and initial probability distributions as estimable parameters, we have a total of $K_T \equiv M(p_T d_T^2 + q_T d_T d_T^* + d_T(d_T+1)/2) + M^2$ parameters. Furthermore, the dimensions of the model, namely $d_T$, $d_T^*$, $p_T$, and $q_T$, are allowed to increase with the sample size subject to regularity conditions. Nonetheless, it is easy to see that, especially with many regimes, such a set-up could easily lead to a high-dimensional problem where the number of parameters exceed the sample size.  

To address this, we assume that the model in \eqref{mainmodel} is sparse. This means that only a small subset of parameters, relative to the sample size, are truly non-zero, while the rest are regarded as irrelevant. Effectively, this means that the 'true' model with only the relevant parameters is a low-dimension system. A key question is thus whether we are able to asymptotically recover the 'true' model given the full data without \textit{a priori} knowledge of the sparsity pattern and the latent states.      

Low-dimensional MS models, where the number of parameters are small and fixed, are ubiquitous in macroeconometrics and empirical finance. Univariate and multivariate MS models have been used in investigations of the business cycle \citep[e.g.][]{hamilton1989new}, monetary policy \citep[e.g.][]{sims2006were}, exchange rates and currency crisis \citep[e.g.][]{cerra2005did, ichiue2011regime}, forecast combinations \citep{elliott2005optimal}, equity returns \citep[e.g.][]{henkel2011time}, and asset allocation \citep[e.g.][]{ang2002international, guidolin2007asset, guidolin2008international} among other applications\footnote{See the surveys by \cite{hamilton2016macroeconomic} and \cite{ang2012regime} for a more extensive list of theoretical and empirical applications in macroeconomics and finance respectively}.

Many empirical investigations in macroeconomics and finance are fundamentally questions about high-dimensional endogenous systems. For example, \cite{bianchi2019modeling} study the stock returns of 83 S\&P100 firms, within the framework of Markov-switching Seemingly Unrelated Regressions, to address questions on network centrality and systemic risk. If we consider larger models or when the number of variables are allowed to diverge, the modeling of such systems may lead to parameter proliferation which causes estimation to be computationally infeasible or unstable. On the other hand, forecasting or policy analysis that restricts the system to only a few variables may suffer from significant omitted variable bias \citep[See for e.g.][]{chan2020reducing, koop2013large, banbura2010large}.

There are at least four traditional responses to this problem: (i) using aggregated data to conserve on degrees of freedom (e.g. stock indices instead of individual prices); (ii) limiting the number of endogenous variables either in an ad-hoc fashion or by invoking economic theory; (iii) factor-augmented approaches introduced by \cite{bernanke2005measuring} and related work on factor analysis by \cite{stock2002forecasting} and \cite{bai2008large}, and (iv) Bayesian estimation when the number of parameters is relatively large but still smaller than the sample size to obtain more stable parameter estimates. Note that (i) and (ii) are compromises and might entail a change in research question, while (iii) does indeed work with high-dimensional data and can be extended to the regime-switching setting as in \cite{liu2016regime}.

More recently, \cite{kock2015oracle} introduced an alternative to factor models in estimating stationary high-dimensional VAR(p) models by means of shrinkage. The authors showed that penalization via the (adaptive) least absolute shrinkage and selection operator (Lasso) allows one to asymptotically recover the sparsity patterns in the coefficient matrices (here, $A_{j,T}$).  \cite{han2015direct} show similar theoretical results under milder conditions but with the number of lags fixed at 1, while \cite{zhu2020nonconcave} provide similar results for the smoothly clipped absolute deviation (SCAD) penalty.  

The model in \eqref{mainmodel} can be viewed as a generalization of \cite{kock2015oracle} by introducing dependency of the multivariate process and its sparsity pattern on a latent state, while allowing for exogenous covariates. We propose two shrinkage type estimators for high-dimensional MS-VARs using either the Lasso or the SCAD penalty, which is a folded concave penalty\footnote{See \cite{fan2014strong} for a characterization.}. We include SCAD penalization in our investigation because it requires only mild conditions for model selection consistency as opposed to the Lasso\footnote{The Lasso requires a strong irrepresentability condition \citep{zhao2006model}, which may not hold in some empirical settings.}. More practically, folded concave penalties have been documented to yield better finite sample performance in empirical applications.

The extension for MS is notably non-trivial as it classically requires some form of MLE, which is not only computationally more intensive compared to the equation-by-equation Lasso framework in \cite{kock2015oracle}, but also theoretically more challenging. Furthermore, the inclusion of folded concave penalties to a optimization problem generally introduces multiple local optima. Existing asymptotic results on high-dimensional penalized MLE, for example \cite{kwon2012large} for the SCAD penalty, establish oracle properties for a theoretic local optimum\footnote{The oracle property is satisfied when the penalized MLE is asymptotically equivalent to the estimator obtained from maximizing the likelihood with irrelevant parameters and penalty terms excluded.}. However, there is no guarantee that a computed solution to our penalized likelihood function is indeed the desired local optimum. For that to be true unequivocally, we require at least an assumption of strict concavity which is only possible when the number of parameters is smaller than the sample size (see theorem 3 of \cite{kim2008smoothly} or theorem 2 of \cite{kwon2012large}), and hence is not directly applicable here. 

To deal with this without deviating from our high-dimensional setting, we adopt a similar estimation strategy as in \cite{fan2014strong}, which relies on the local linear approximation (LLA) algorithm for the folded concave penalties introduced by \cite{zou2008one}. This approach does not attempt to find the desired local optimum, but instead seeks a lower probability bound to the claim that the computed solution is indeed the desired optimum.
As the SCAD penalty requires an initial estimator, theorem 1 of \cite{fan2014strong} claims that if we can find an initial estimate that is asymptotically close to the true parameter vector, in a sense to be made clear in \thref{fsconst}, then the LLA algorithm delivers the oracle estimator in a single step with probability approaching one. Borrowing from their terminology, we will call such an initial estimator "localizable". However, given that the likelihood function of MS models tend to be general and potentially multimodal, we discipline our investigation by requiring the likelihood surface at the true parameter to be locally concave and the oracle solution to be unique. Such a restriction is nonetheless considerably weaker than the strict concavity condition required in \cite{kim2008smoothly} or \cite{kwon2012large}. Taken together, this suggests a two-stage approach for the SCAD estimator, similar to that in \cite{li2015model} and \cite{chen2020time}, with the search of a localizable initial estimator in the first stage, and the SCAD-penalized MLE subsequently.

Our results open up the possibility of studying Markov regime-switching dynamics in high-dimensional macroeconomic and financial systems. To illustrate, we extend the investigation on short-horizon stock return predictability considered in \cite{henkel2011time} to incorporate 14 aggregate return predictors from \cite{welch2008comprehensive}, up from the original 4, which mitigates potential omitted variable bias. This results in a relatively high-dimensional system with 711 estimable parameters. Penalized maximum likelihood estimation with the SCAD penalty yields the often reported counter-cyclicality in return predictability. Furthermore, the variable selection property of our estimators helps to identify predictors that contribute strongly to predictability during economic contractions but are otherwise irrelevant in expansions.

The rest of this paper is organized as follows. Section 2 describes the penalized maximum likelihood problems while Section 3 provides an EM algorithm for the special case of Gaussian errors. Section 4 contains the key asymptotic results. Section 5 consider Monte Carlo experiments to assess the finite sample properties of the proposed estimators, while section 6 applies it to the problem of short-horizon return predictability. Finally, section 7 concludes. All proofs are collected in the appendix.

\section{Problem formulation}\label{problem}

In this section, we describe the maximum likelihood problem in a manner amenable to the derivation of theoretical results. This construction is however, not convenient for computation, which will be accomplished in section \ref{algorithm}. There, we will rely on the Expectation-Maximization (EM) algorithm.

\subsection{Notation and sparsity}\label{notation}

Before proceeding, we define some notation. For this section and the next, we suppress the dependence of the parameters on the sample size for notational convenience.

To begin, set $\phi \equiv (\theta_A^\top, \theta_B^\top, Q^{ND\top}, Q^{D\top}, \pi^\top)^\top$ to include all parameters across all states. Let $A(s) = [A_1(s), \ldots, A_{p_T}(s)]$, and $A = [A(1), \ldots, A(M)]$, then $\theta_A = vech(A)$, which is a $M p_T d_T^2 \times 1$ vector. $\theta_B$ is a $Mq_Td_Td_T^* \times 1$ vector that is defined analogously. Recall that the precision matrix is given as $Q(s) = \Sigma(s)^{-1}$. Let $Q^{ND}$ be the vector containing unique off-diagonal elements of the precision matrix across all states i.e.  $Q^{ND} = (q_{21}(1), q_{31}(1), \ldots, q_{d_T1}(1), q_{23}(1), \ldots, q_{d_T(d_T-1)}(1), q_{21}(2),  \ldots,  q_{d_T(d_T-1)}(M))^\top$, and $Q^{D}$ includes the diagonal elements: $Q^{D} =(q_{11}(1),\ldots,q_{d_Td_T}(1),\ldots, q_{d_Td_T}(M))^\top$.  $Q = (Q^{ND\top}, Q^{D\top})^\top$ is $M d_T (d_T +1)/2 \times 1$. Finally, $\pi = (p_{1 \shortrightarrow 1}, p_{1 \shortrightarrow 2}, \ldots, p_{1 \shortrightarrow M},\ldots, p_{M \shortrightarrow M}, p_1, \ldots, p_M)^\top$ is a vector of transition probabilities and initial distributions.

Written this way, it is clear that our characterization of sparsity is equivalent to saying that the vectors $\theta_A$ , $\theta_B$, and $Q^{ND}$ are sparse. Formally, we assume that only $h^A$, $h^B$, and $h^{Qnd}$ elements of $\theta_A$, $\theta_B$, and $Q^{ND}$ respectively, are non-zero and that $h^A + h^B + h^{Qnd} \ll T$. Sparsity cannot be imposed on $Q^{D}$ to maintain positive definiteness. 

\subsection{Two estimators of a high-dimensional MS-VAR}


To begin, let $\mathcal{I}^{t-1}_{t-v} = (y_{t-v}, \ldots, y_{t-1}, x_{t-v}, \ldots, x_{t-1} )$ be the information set from time $t-v$ to $t-1$ for some integer $v$, $\Phi_{s} = (A(s), B(s), Q(s))$ be the state-specific VAR parameters, and define $\mathcal{Y}_T = (y_1, \ldots, y_T)$ and analogously for $\mathcal{X}_T$. Without loss of generality, assume that $p_T = \max\{p_T, q_T\}$ is the largest lag, and thus we can denote the conditional density of $y_t$ from \eqref{mainmodel} as $g(y_t| \mathcal{I}^{t-1}_{t-p_T}; \Phi_{S_t})$.  

Then, it can be shown that the (non-penalized) conditional likelihood is given as 
\begin{equation}\label{likelihood}
\mathcal{L}(\mathcal{Y}_T| \mathcal{X}_T; \phi) = \sum_{S_{p_T+1}=1}^M \ldots \sum_{S_T=1}^M p_{S_{p_T+1}}(\phi) \prod_{t = p_T+2}^T p_{(S_{t-1}) \shortrightarrow (S_t)}(\phi) \prod_{t=p_T+1}^T g(y_t|\mathcal{I}^{t-1}_{t-p_T};\Phi_{S_t}).   
\end{equation}
Here we emphasize that $p_{i \shortrightarrow j}$ and the initial distributions $p_j$ are functions of the parameter vector $\phi$. As prefaced earlier, direct optimization of \eqref{likelihood} is challenging because the transition probabilities $p_{i \shortrightarrow j}$ are highly non-linear functions of the parameters. To deal with this issue, we will rely on a modified EM algorithm in section \ref{algorithm}.

\subsubsection{Lasso \& gLasso}

We propose solving the following optimization problem with Lasso penalty for VAR coefficients and graphical Lasso (gLasso) penalty for the precision matrix \citep{friedman2008sparse} to induce sparsity,
\begin{equation} \label{fstage}
\max_\phi \ T^{-1}\log \mathcal{L}(\mathcal{Y}_T|\mathcal{X}_T; \phi) - \sum_{s=1}^{M}\bigg\{ \lambda^{Lasso}\bigg(\sum_{m=1}^{d_T} \sum_{n=1}^{d_Tp_T} |a_{mn}(s)|  + \sum_{m=1}^{d_T^*} \sum_{n=1}^{d_T^*q_T}|b_{mn}(s)|\bigg)  + \lambda^{gLasso} \sum_{m \neq n} |q_{mn}(s)| \bigg\} 
\end{equation}
where $a_{mn}(s)$ refers to the element in the $m^{th}$ row and $n^{th}$ column of $A(s)$, and analogously for $b_{mn}(s)$ and $q_{mn}(s)$ for $B(s)$ and $Q(s)$ respectively. $\lambda^{Lasso}$ and $\lambda^{gLasso}$ are penalty terms. Note that only the non-diagonal elements are penalized in the sparse precision matrix estimation. It is possible to set $\lambda^{Lasso}$ to be different for parameters in $A(s)$, and $B(s)$ as long as the penalties are proportional to one another, however, doing so may introduce significant computational costs when tuning the penalty terms. Nonetheless, we call the solution to \eqref{fstage} the Lasso estimator.

\subsubsection{SCAD}

To use the LLA \citep{zou2008one} for the SCAD problem, we require a localizable initial estimate, $\tilde{\phi}$. As we show later in section \ref{asymptotic}, the Lasso estimate above satisfies this property. Hence, we re-optimize the log-likelihood with the Lasso estimate as $\tilde{\phi}$ with the LLA for the SCAD penalty. This utilizes the first derivative of the penalty which is defined as
\begin{equation}
p_\lambda^{'} (x) = \lambda\bigg[ 1(x\leq \lambda) + \frac{(a\lambda-x)_+}{(a-1)\lambda}1(x>\lambda)\bigg], \label{SCADpenalty}
\end{equation} 
where $(m)_+ = m$ if $m > 0$ and $0$ otherwise, $\lambda$ is a penalty parameter, and $a>2$ is a constant. Here, we set $a=3.7$ as suggested in \cite{fan2001variable}. The optimization problem is given by
\begin{align} \label{sstage}
\max_{\phi} \ T^{-1}\log \mathcal{L}(\mathcal{Y}_T| \mathcal{X}_T; \phi) - \sum_{s=1}^M \bigg\{  \sum_{m=1}^{d_T}&\sum_{n=1}^{d_Tp_T} p_{\lambda}^{'}(|\tilde{a}_{mn}(s)|)|a_{mn}(s)| + \notag \\ &\sum_{m=1}^{d_T^*}\sum_{n=1}^{d_T^*q_T} p_{\lambda}^{'}(|\tilde{b}_{mn}(s)|)|b_{mn}(s)| +   \sum_{m \neq n}  p_{\lambda^*}^{'}(|\tilde{q}_{mn}(s)|)|q_{mn}(s)| \bigg\}
\end{align}
where $\lambda$ and $\lambda^*$ are penalty terms, and parameters with tilde are from the initial estimate $\tilde{\phi}$.

A key theoretical advantage of using the SCAD penalty instead of simply stopping once we have obtained the Lasso estimate, is that we can, under relatively mild conditions, attain selection consistency, or in other words, asymptotically recover the true sparse support of the parameters. On the other hand, penalization with Lasso requires strong irrepresentability conditions \citep{zhao2006model}. However, if prediction is the main goal of estimation, then both Lasso and SCAD are applicable. 

\section{EM algorithm for Gaussian errors}\label{algorithm}

In this section, we propose an EM algorithm to solve \eqref{fstage} and \eqref{sstage} similar to that of \cite{monbet2017sparse}. For concreteness and parsimony, we focus only on the scenario where $\varepsilon_t \sim^{i.i.d} N(0, \Sigma(S_t))$ in this section, which implies that the conditional density $g(y|\cdot)$ is Gaussian. Not only is this assumption standard in the VAR literature, it also induces a closed form problem which simplifies the estimation of the VAR parameters.

The EM algorithm was proposed to estimate models with incomplete or hidden data. \cite{baum1970maximization} applied the algorithm to estimate hidden Markov models, which is a general class that encompasses many MS models in econometrics. Intuitively, the algorithm works not by directly optimizing the likelihood function in \eqref{likelihood}, which as mentioned earlier is a highly non-linear function of the parameters, but instead optimizes a constructed auxiliary function (label it $\Omega(\cdot)$) to derive a monotonically increasing lower bound on the value of the original likelihood.  

Formally, the algorithm works iteratively with two steps per iteration: expectation (E) and maximization (M). With each iteration applied to $\Omega(\cdot)$, we obtain updates on the parameters. The corresponding likelihood $\mathcal{L}(\mathcal{Y}_T|\mathcal{X}_T;\phi)$ is guaranteed to be non-decreasing with each update and eventually finds a stationary point\footnote{This can mean a local or global maximum, or a saddle point.} in the likelihood surface \citep{dempster1977maximum}. Although the statistical guarantees on the EM algorithm have been developed for the low-dimensional context, it is not difficult to show that Theorem 1 of \cite{dempster1977maximum} (monotonicity of the algorithm) will still hold with penalization. 

Since both estimators in section \ref{problem} entail a maximization problem, we can apply the EM algorithm to optimize either \eqref{fstage} or \eqref{sstage}. The E-step for both problems will be similar, while the M-step is different because of differences in penalization. 

To begin, we define the auxiliary function in the E-step.

\textbf{E-step}. We consider the complete-data log-likelihood given as $\ell(\mathcal{S}_T,\mathcal{Y}_T;\mathcal{X}_T, \phi)$ where $\mathcal{S}_T = (S_1, \ldots, S_T)$. Note that the incomplete-data likelihood in \eqref{likelihood} can be written as
\begin{equation*}
\mathcal{L}(\mathcal{Y}_T|\mathcal{X}_T; \phi) = \int \exp(\ell(\mathcal{S}_T,\mathcal{Y}_T| \mathcal{X}_T;  \phi))d \mathcal{S}_T,
\end{equation*}
and thus
\begin{equation*}
\ell(\mathcal{S}_T,\mathcal{Y}_T;\mathcal{X}_T, \phi) = \log \bigg[ p_{S_{p_T+1}}(\phi) \prod_{t = p_T+2}^T p_{(S_{t-1}) \shortrightarrow (S_t)}(\phi) \prod_{t=p_T+1}^T g(y_t|\mathcal{I}^{t-1}_{t-p_T};\Phi_{S_t}) \bigg].
\end{equation*}
The log-likelihood is called 'complete' because it treats the latent state as observable data. However, since we do not actually observe it, we consider its conditional expectation, $E[\ell(\mathcal{S}_T,\mathcal{Y}_T;\mathcal{X}_T, \phi) \\| \mathcal{I}_T, \phi^{(j-1)}]$, where $\phi^{(j-1)}$ is the estimate from the previous $(j-1)^{th}$ EM iteration and is treated as pre-determined during the current iteration. Since we have additive penalties, we can construct the auxiliary function as
\begin{equation} \label{omega}
\Omega^{\mathrm{N}}(\phi, \phi^{(j-1)}) \equiv E[\ell(\mathcal{S}_T,\mathcal{Y}_T;\mathcal{X}_T, \phi)| \mathcal{I}_T, \phi^{(j-1)}] - \text{pen}(\mathrm{N})
\end{equation}
where pen(N) for N $\in \{\text{Lasso, SCAD}\}$ refers to either group of penalties. Ignoring penalties for now, we have
\begin{align*}
E[\ell(\mathcal{S}_T,\mathcal{Y}_T;\mathcal{X}_T, \phi)&| \mathcal{I}_T, \phi^{(j-1)}] = \sum_{s=1}^{M} \sum_{t=p_T+1}^T P(S_t = s|\mathcal{I}_T; \phi^{(j-1)}) \log g(y_t|\mathcal{I}^{t-1}_{t-p_T}; \Phi_s) \notag \\
&+ \sum_{s,s^{'}=1}^M \sum_{t=p_T+2}^T P(S_t=s^{'},S_{t-1}=s|\mathcal{I}_T, \phi^{(j-1)}) \log P(S_t = s^{'} | S_{t-1} = s, \mathcal{I}^{t-1}_{t-p_T}; \pi) \\
&+ \sum_{s=1}^M P(S_{P_T+1}=s| \mathcal{I}_{t-p_T}^{t-1}; \pi) \\
&\equiv \sum_{s=1}^M Z_1(s; \Phi_s) + \sum_{s,s^{'}=1}^M Z_2(s,s^{'}; \pi) + \sum_{s=1}^M Z_3(s; \pi).
\end{align*}
By construction, pen(N) would be relevant to the optimization of VAR parameters in $Z_1(s)$. On the other hand, $Z_2(s,s^{'})$ and $Z_3(s)$ are only a function of $\pi$ and thus can be maximized by standard optimization procedures as this is a low-dimensional problem. 

Since we have Gaussianity of $g(y_t|\cdot)$, we can show that
\begin{equation}
\argmax_{(A(s),B(s),Q(s))} Z_1(s; \Phi_s) = \argmax_{(A(s),B(s),Q(s))} \log |Q(s)| - \Tr{(\hat{S}(s) Q(s))}, 
\end{equation}
where
\begin{equation*}
\hat{S}(s) = \frac{\sum_{t=p_T+1}^T P(S_t = s|\mathcal{I}_T; \phi^{(j-1)}) \times \omega_t \omega_t^\top}{\sum_{t=p_T+1}^T P(S_t = s|\mathcal{I}_T; \phi^{(j-1)})},
\end{equation*}
and 
\begin{equation*}
\omega_t = y_t - \sum_{j=1}^{p_T}A_{j}(S_t)y_{t-j} - \sum_{j=1}^{q_T} B_{j}(S_t)x_{t-j}.
\end{equation*}

Note that $Z_2(s,s^{'}; \pi)$ contains the smoothed probabilities $P(S_t=s^{'}|S_{t-1}=s, \mathcal{I}_T, \phi^{(j-1)})$, which can be computed via an iterative backward-forward recursion (see for e.g. \cite{hamilton1990analysis}). 

\textbf{M-step}. Here, we maximize the auxiliary function in \eqref{omega} with respect to $\phi$. Suppose we are executing the EM algorithm for the N estimator (i.e. either \eqref{fstage} or \eqref{sstage}). Then, for all states $s$, we solve
\begin{equation}\label{firstpart}
\argmax_{(A(s),B(s),Q(s))} \log |Q(s)| - \Tr{(\hat{S}(s) Q(s))} - \text{pen(N},s\text{)},
\end{equation}
\begin{equation}\label{secondpart}
\argmax_{\pi} Z_2 (s,s^{'}; \pi) \text{ and } \argmax_{\pi} Z_3 (s; \pi),
\end{equation}
where pen(N,$s$) refers to the $s^{th}$ state penalties for the N estimator. For computational efficiency, it might be convenient to further separate \eqref{firstpart} into two parts. To illustrate, consider a M-step for the Lasso. Firstly, fix the coefficient values $A(s)$ and $B(s)$, and estimate
\begin{equation*}
\argmax_{Q(s)} \log|Q(s)| - \Tr (\hat{S}(s)Q(s)) - \lambda^{gLasso} \sum_{m \neq n} |q_{mn}(s)|.
\end{equation*}
Next, given an estimate of $Q(s)$, we optimize
\begin{equation*}
\argmax_{A(s), B(s)} - \Tr (\hat{S}(s)Q(s)) - \lambda^{Lasso} \bigg( \sum_{m=1}^{d_T} \sum_{n=1}^{d_T p_T}|a_{mn}(s)| + \sum_{m=1}^{d_T^*} \sum_{n=1}^{d_T^* q_T} |b_{mn}(s)| \bigg).
\end{equation*}
This sequential partitioning helps with computation because we can now individually apply fast (block) coordinate descent algorithms \citep{friedman2007pathwise, friedman2008sparse} to each part of the problem. Note that these algorithms will also work with the SCAD penalty as defined in \eqref{sstage}.

The M-step provides an update of the parameters $\phi^{(j)}$. It can be shown that repeated updating of $\phi^{(j)}$ will lead the respective original likelihoods in \eqref{fstage} or \eqref{sstage} to either increase or remain constant in value, but not decrease. Hence, the maximizer can be found subject to some termination condition on the implied increments of the original penalized likelihood. 

To implement the algorithm, we require an initial vector of parameters $\phi^{(0)}$. We recommend using reasonable randomly generated values for the Lasso problem, and subsequently the computed Lasso optimum as $\phi^{(0)}$ for the SCAD problem.

\subsection{Selecting tuning parameters}

Similar to \cite{kock2015oracle}, we propose selecting $\lambda^{Lasso}$ and  $\lambda^{gLasso}$ for the Lasso estimator, and $\lambda$ and $\lambda^*$ for the SCAD estimator, using a modified version of the Bayes Information Criterion (BIC). Specifically, for either Lasso or SCAD, choose $\lambda_1$ and $\lambda_2$ to minimize 
\begin{equation*}
BIC(\lambda_1,\lambda_2) = \log(\text{pooled}_{SSR}) + C_T (l \log(T)/ T)
\end{equation*}
where
\begin{equation*}
\text{pooled}_{SSR} = \left| \sum_{t={p_T+1}}^T\sum_{m=1}^M \hat{P}(S_t = m|\mathcal{I}_T; \hat{\phi}) \left(y_t - \sum_{j=1}^{p_T}\hat{A}_{j,T}(m)y_{t-j} + \sum_{j=1}^{q_T} \hat{B}_{j,T}(m)x_{t-j}\right)  \right|,
\end{equation*}
and parameters with 'hats' indicate estimates. Following \cite{wang2009shrinkagetuning}, $C_T$ is set at $\log K_T$, where we recall that $K_T$ is the total number of parameters in the system, to help obtain consistency of the BIC in high-dimensional regressions, while $l$ is the number of estimates that are identified as non-zero in the system. Simulation results in section \ref{sim} indicate that tuning parameter selection with BIC can yield consistent estimates.

\section{Asymptotic theory}\label{asymptotic}

We establish theoretical properties for each estimator in a double asymptotic framework, where we allow the sample size $T \rightarrow \infty$ and the dimensions of the candidate models $p_T, q_T, d_T,$ and $d_T^*$ to diverge at appropriate rates. By extension, it is natural to allow the dimensions of the true non-zero parameters $h^A_T$, $h^B_T$, and $h^{Qnd}_T$ to grow albeit subject to stricter restrictions. Consequently, the dimension of the true parameter vector, $\phi_T$, may extend to infinity.

Traditionally, theoretical results on consistency in MS models with finite state spaces rely on limit theorems from random matrix theory to attain some generalized form of the Kullback-Leibler divergence \citep[for e.g.][]{leroux1992maximum, francq1998ergodicity}, whereby consistency follows from an identification condition. This strategy is challenging in the high-dimensional context because of the dependence of $\phi_T$ on the sample size. Instead, we adopt a modified approach to \cite{fan2004nonconcave} to establish consistency and oracle properties under a diverging parameter framework.

Recall that $K_T$ is the total number of VAR and transition matrix parameters from \eqref{mainmodel}. Let $\phi^*_T \in \Theta_T$ be a vector of true (sparse) parameters where $\Theta_T \subset \mathbb{R}^{K_T}$ is an open subset. In addition, let $K^{Sp}_T = h_T^A + h_T^B + h_T^{Qnd}$ be the number of non-zero parameters out of those that are subject to penalization (i.e. the VAR parameters), and let $K_T^* = K^{Sp}_T  + Md_T + M^2$ be the total number of non-zero parameters in the true parameter vector where $M^2$ and $Md_T$ are from the transition matrix and the diagonals of variance-covariance matrices respectively. Without loss of generality, we assume that $\phi^*_T$ can be re-arranged in the following form
\begin{equation}\label{phistar}
\phi_T^* = (\underbrace{\phi_1^*, \ldots, \phi_{K_T^{Sp}}^*}_{\substack{K_T^{Sp} \text{ non-zero} \\ \text{VAR parameters}}}, \underbrace{0, \ldots, 0}_{\substack{\text{irrelevant} \\ \text{VAR parameters}}}, \underbrace{q_{11}^*(1), \ldots, q_{d_Td_T}^*(M)}_{\substack{ Md_T \text{ parameters from} \\ \text{diagonals of precision matrices}}}, \underbrace{p_{1\shortrightarrow 1}^*, \ldots, p_{M \shortrightarrow M}^*}_{\text{transition probabilities}} )^\top.
\end{equation}

\subsection{Regularity conditions}

We impose the following regularity conditions for deriving our asymptotic results. Let $\mathcal{L}(\mathcal{Y}_T| \mathcal{X}_T; \phi_T) = \mathcal{L}_T(\phi_T)$. In addition, $\nabla^{k}$ represents the $k^{th}$ derivative with respect to $\phi_T$, and $\nabla^{k}_{j_1, \ldots, j_k}$ represent the $k^{th}$ derivative with respect to the $j_1, \ldots, j_k^{th}$ element in $\phi_T$.  

\begin{itemize}
\setlength{\itemindent}{1em}
\item[\mylabel{itm:stat}{(A1)}] For all $\phi_T \in \Theta_T$, $\{(y_t, x_t)\}_{t=0}^\infty$ is a stationary and ergodic process. The Markov chain $\{S_t\}$ is irreducible and aperiodic.

\item[\mylabel{itm:cont}{(A2)}] For all $i, j$, the functions $p_j(\cdot)$ and $p_{i \shortrightarrow j}(\cdot)$ are twice continuously differentiable over $\Theta_T$. Furthermore, given $I_{t-1}$ and $S_t$, $g(y_t|\mathcal{I}^{t-1}_{t-v}; \Phi_{S_t,T})$ is a probability density function with two continuous derivatives over $\Theta_T$ for any integer $v \leq t$. 

\item[\mylabel{itm:momentcondition}{(A3)}] For all $S_t$, $j,k,l$, we have: \\ (i) $E[\sup_{\phi_T \in \Theta_T} |\nabla^{1}_j \log g(y_t|\mathcal{I}_{t-v}^{t-1}; \Phi_{S_t})|^2] < \infty$ and $E[\sup_{\phi_T \in \Theta_T} |\nabla^{2}_{j,k} \log g(y_t|\mathcal{I}_{t-v}^{t-1}; \Phi_{S_t})|^2] < \infty$; (ii) $E[\sup_{\phi_T \in \tilde{\Theta}_T} |\nabla^{3}_{j,k,l} \log g(y_t|\mathcal{I}_{t-v}^{t-1}; \Phi_{S_t})|^2] < \infty$, where $\tilde{\Theta}_T$ is defined in \ref{itm:concave} and \ref{itm:third}(i). 

\item[\mylabel{itm:minorization}{(A4)}] Let $\rho(y_t) = \sup_{\phi_T \in \Theta_T} \max_{s,s^{'} \in \{1, \ldots, M\}} \frac{g(y_t|\mathcal{I}_{t-v}^{t-1}; \Phi_{s})}{g(y_t|\mathcal{I}_{t-v}^{t-1}; \Phi_{s^{'}})}$, and assume that $P(\rho(y_t) = \infty | \mathcal{I}_{t-v}^{t-1}, S_t = s) < 1$ for all $s \in \{1 \ldots, M\}$.  

\item[\mylabel{itm:concave}{(A5)}] (i) Assume that there exists an open subset $\tilde{\Theta}_T \subset \Theta_T$ such that $\tilde{\phi}_T,\phi_T^* \in \tilde{\Theta}_T$, and $\log \mathcal{L}_T(\phi_T)$ is locally concave over $\tilde{\Theta}_T$. 

(ii) Let $\boldsymbol{u}_1, \boldsymbol{u}_2, \boldsymbol{u}_3, \boldsymbol{v}_1$, and $\boldsymbol{v}_2$ be vectors that share the same dimensions as $\theta_A, \theta_B, Q^{ND}, Q^{D}$ and $\pi$ respectively. Fix a scalar constant $W$ that can be sufficiently large, and define the set $\Omega(W) = \{ \boldsymbol{u} = (\boldsymbol{u}_1^\top, \boldsymbol{u}_2^\top, \boldsymbol{u}_3^\top, \boldsymbol{v}_1^\top, \boldsymbol{v}_2^\top)^\top | \ \|\boldsymbol{u}\| = W;\ \|u_1\|_1 + \|u_2\|_1 + \|u_3\|_1 \leq C [\sum_{i \in S(h_T^A)} |u_{1i}| + \sum_{i \in S(h_T^B)} |u_{2i}| + \sum_{i \in S(h_T^{Qnd})} |u_{3i}|] \}$, where $\|\cdot\|_1$ refers to the sum of absolute values of vector elements, and $C > 1$ is some constant. Let $\gamma_T$ be some scalar function that depends on $T$, then define the ball $\tilde{\Theta}_{T,\gamma,W} = \{(\phi_T^* + \gamma_T\boldsymbol{u})| \boldsymbol{u} \in \Omega(W)\}$. Assume that $\tilde{\Theta}_{T,\gamma,W} \subseteq \tilde{\Theta}_T$.

\item[\mylabel{itm:third}{(A6)}] (i) $\log \mathcal{L}_T (\phi_T)$ admits a third derivative over $\tilde{\Theta}_T$ which includes $\phi^*_T$; (ii) There exists an open set $\tilde{\Theta}^*_T \subseteq \tilde{\Theta}_T$ and $\phi^*_T\in \tilde{\Theta}^*_T$ for which $\nabla^2 \log \mathcal{L}_T (\phi_T)$ is concave.

\item[\mylabel{itm:eigen}{(A7)}] (i) Let the information matrix be $I_T (\phi_T^*) = E[(\nabla^{1} \log \mathcal{L}_T(\phi_T^*))(\nabla^{1} \log \mathcal{L}_T(\phi_T^*))^\top]$, and let $I_T^{(\neg 0)}(\phi_T^*)$ refer to the non-zero submatrix constructed from $I_T (\phi_T^*)$. Assume that $I_T^{(\neg 0)}(\phi_T^*)$ is positive definite. 

(ii) In addition, assume that
\begin{equation*}
0 < \rho_1 \leq \inf_{\omega \in \Omega(W)} \omega^\top(-\nabla^2 \log \mathcal{L}_T(\phi_T^*)) \omega \leq \sup_{\omega \in \Omega(W)} \omega^\top(-\nabla^2 \log \mathcal{L}_T(\phi_T^*))\omega \leq \rho_2 < \infty,
\end{equation*}
with probability approaching one.

\item[\mylabel{itm:rates}{(A8)}] (i) $K_T^* = o(T^{1/4})$; (ii) The Lasso penalty terms satisfy $\lambda^{Lasso} \propto \lambda^{gLasso}$, where $\lambda^{Lasso} \rightarrow 0$ satisfies $(\sqrt{T}\lambda^{Lasso})^{-1} \rightarrow 0$. (iii) The SCAD penalty terms satisfy $\lambda \propto \lambda^*$, where $\lambda \rightarrow 0$, and $[\sqrt{K_T^*}\lambda^{Lasso}]\lambda^{-1} \rightarrow 0$. Furthermore, $\frac{K_T}{T \lambda^2} \rightarrow 0$, where recall that $K_T$ is the number of candidate parameters.

\item[\mylabel{itm:nonzero}{(A9)}] Assume $\min_{1\leq j \leq K_T^{Sp}} |\phi_i^*| > \lambda$ such that $\min_{1\leq j \leq K_T^{Sp}} |\phi_i^*|/\lambda \rightarrow \infty$, where $\phi_i^*$ are true non-zero parameters as described in \eqref{phistar}. 

\end{itemize}

\noindent \textbf{Remarks.} \ref{itm:stat}-\ref{itm:minorization} are standard assumptions in the MS literature on asymptotic normality \cite[see ][]{bickel1998asymptotic}. The assumption of local concavity in \ref{itm:concave} is much weaker than that of strict concavity commonly assumed in high-dimensional regularized MLE problems. We need the likelihood to be (locally) concave between the localizable initial estimator and the true parameter to apply theorem 1 of \cite{fan2014strong} so that the SCAD estimation initialized with Lasso estimates delivers the oracle result. The assumption of a locally concave neighborhood is not uncommon in the theoretical literature on high-dimensional problems. For example, in the context of a two-step estimation procedure for high-dimensional sparse principal components analysis, \cite{jankova2018biased} assumes that a rough initial estimate can be found in a locally convex neighborhood of the population parameter. \ref{itm:third}(i) is commonly assumed in the SCAD literature and helps us to establish selection consistency, while \ref{itm:third}(ii) is essentially identical to assumption 2(ii) in \cite{li2015model}. \ref{itm:eigen}(ii) is a modification of the restricted eigenvalue condition in \cite{bickel2009simultaneous}, and this particular formulation follows assumption 2(iii) in \cite{li2015model}. \ref{itm:rates} restricts the number of true parameters $K_T^*$ to only increase at a rate slower than $T^{1/4}$. Furthermore, $\lambda^{lasso} \rightarrow 0$ but converges slower than $1/\sqrt{T}$, and $\lambda \rightarrow 0$ but slower than $\sqrt{K_T^*}\lambda^{lasso}$.  \ref{itm:nonzero} is a standard assumption on the minimum signal strength of relevant parameters. This condition is required for proving the oracle property of the SCAD estimator and mirrors the characterization in \cite{zhu2020nonconcave}.

\subsection{Asymptotic results}

Our first result states that the Lasso estimator is consistent. Let $\|\cdot\|$ denote the $\ell_2$ norm.

\begin{proposition}\thlabel{fsconst}
Under the conditions of \ref{itm:stat}-\ref{itm:rates}, the Lasso estimate $\tilde{\phi}_T$ satisfies
\begin{equation}
\|\tilde{\phi}_T - \phi_T^* \| = O_p(\sqrt{K_T^*} \lambda^{Lasso}) = o_p(1). \label{lassobound}
\end{equation}
\end{proposition}

We obtain the final equality because of (A8). This means that the Lasso estimator exhibits estimation consistency, although we do not guarantee that it is able to asymptotically identify the relevant parameters. 

Next, define $\hat{S}=\{j: |\hat{\phi}_{T,j}| > 0\}$ where $\hat{\phi}_{T,j}$ is the $j^{th}$ element in $\hat{\phi}_{T}$. Hence, $\hat{S}$ is the index set of estimated non-zero parameters from the SCAD procedure. Define $S_0$ for $\phi_T^*$ in an analogous manner (i.e. the index set of truly relevant population parameters). The following result shows that the SCAD estimator, initialized by the Lasso estimates, achieves not just estimation consistency, but also selection consistency.

\begin{theorem}\thlabel{consistency}
Let $\hat{\phi}_T$ be the MLE to the SCAD problem in \eqref{sstage} initialized with Lasso estimates. Under the conditions of \ref{itm:stat}-\ref{itm:nonzero}, we have that
\begin{itemize}
\setlength{\itemindent}{1em}
\item[(1)] $\|\hat{\phi}_T - \phi_T^*\| = O_p(\sqrt{K_T^*/T}) = o_p(1)$,
\item[(2)] $P(\hat{S} = S_0) \rightarrow 1$.
\end{itemize}   
\end{theorem}      

Part (1) of \thref{consistency} states that the SCAD estimator is consistent since $\sqrt{K_T^*/T} \rightarrow 0$ by (A8). The second part claims that the sparsity property holds. In other words, we are able to exactly distinguish the parameters that are truly non-zero from those that are irrelevant in the theoretical limit. Note that these results hold even if the dimensions of the parameter vectors diverge.  

Next, we discuss the asymptotic normality of the SCAD estimator. To do so, define $\hat{\phi_T}^{(\neg 0)}$ to be the estimate $\hat{\phi}_T$ with the zeroes removed, and similarly for $\phi_T^{*(\neg0)}$. 

\begin{theorem}\thlabel{norm} Assume that conditions \ref{itm:stat}-\ref{itm:nonzero} are satisfied. Then, 
\begin{equation}\label{asymptoticnorm}
\sqrt{T} G_T [I_T^{(\neg 0)}(\phi_T^*)]^{1/2}(\hat{\phi_T}^{(\neg 0)} - \phi_T^{*(\neg0)}) \rightarrow^{d} \mathcal{N}(0,G),
\end{equation}
where $G_T$ is a conformable matrix such that $G_T G_T^\top \rightarrow G$ for positive definite $G$, and $I_T^{(\neg 0)}$ is defined in \ref{itm:eigen}.
\end{theorem}

We remark that the pre-multiplication of $G_T$ helps with the exposition since $\phi_T^*$ can be diverging in dimension. 
\thref{norm} implies that it is asymptotically justifiable to apply the same statistical inference for the estimate obtained from maximizing the problem with \textit{a priori} knowledge on the sparsity pattern, to the SCAD-penalized solution $\hat{\phi}_T$. Here, we note that this convergence in distribution is a pointwise result, and is not guaranteed to hold uniformly with respect to the parameter vector \citep{leeb2005model}. Uniform inference for model selection in multivariate time series is however still a nascent area of research \citep{masini2020machine} and is beyond the scope of this paper.

\section{Monte Carlo Simulation}\label{sim}

This section studies the finite sample properties of the proposed EM algorithm in estimating large MS-VARs using both Lasso and SCAD penalization schemes\footnote{For SCAD, we initialize the algorithm with Lasso estimates as described earlier.}. The number of endogenous variables considered are $d = 10$ and $16$. We consider three experiments\footnote{The numerical experiments here are similar to those considered in \cite{kock2015oracle}.} with 2 states ($M=2$) throughout:
\begin{itemize}
    \item Experiment 1: The data generating process (DGP) is a MS-VAR(1) with the following coefficient matrices. In state 1, $A_{1}(1) = diag(0.8, \ldots, 0.8)$, and $A_{1}(2) = -A_{1}(1)$ for state 2. Effectively, within each state, we have a stationary AR(1) process, as the lagged terms of other variables do not appear in the DGP for a given variable. 
    \item Experiment 2: $A_{1}(1)$ is a block diagonal matrix with upper-left and lower-right non-zero blocks. Each block has dimension $d/2 \times d/2$, and is a tridiagonal matrix with $0.5$ on the diagonal, while the sub- and superdiagonal are set at $-0.45$. All other elements in $A_{1}(1)$ are 0, while $A_{1}(2) = -A_{1}(1)$. The variance-covariance matrices are also non-sparse: $\Sigma(1)_{ij} = 0.7^{|i-j|}$ and $\Sigma(2)_{ij} = 0.4^{|i-j|}$. 
    \item Experiment 3: The DGP is a MS-VAR(2). $A_{1}(1)$ and $A_{1}(2)$ are the same from the second experiment, while $A_{2}(1)_{ij} = (A_{1}(1)_{ij})^2$ and $A_{2}(2) = -A_{2}(1)$. We set $\Sigma(1) = diag(0.8,\ldots, 0.8)$ and $\Sigma(2) = diag(0.4,\ldots,0.4)$. Such a process might be of interest in macroeconomics where the influence of variables in the past is usually weaker than that of recent lags. 
    
\end{itemize}
In all experiments, we exclude an intercept and set the transition probabilities to be $p_{1\rightarrow 1} = p_{2\rightarrow 2} = 0.8$.

500 iterations are generated for sample sizes $T=100,200,$ and $300$.
To evaluate our procedures, we consider the following metrics. The first three metrics are concerned with selection consistency. \textit{True model included} looks at the share of iterations in which the estimate includes the true model for both states (i.e. truly non-zero coefficients are estimated as non-zero). \textit{Selected variables} is the number of non-zero parameters estimated by the system. As it may be the case that the estimation sets a truly non-zero coefficient to zero, it is informative to study the \textit{share of truly non-zero parameters} that are identified correctly as non-zero by the algorithm. Subsequently, we consider estimation consistency as measured by the root mean squared error (\textit{RMSE}) of the parameters. This is given by $\sqrt{\frac{1}{200} \sum_{i=1}^{200} \| \hat{\phi}_T(i) - \phi_T^* \|}$ where $\hat{\phi}_T(i)$ is the estimated parameter vector for iteration $i$ containing all parameters in the system. $RMSE_{VAR}, RMSE_{COV},$ and $RMSE_{p}$ are similarly defined \textit{RMSE} measures for estimated VAR coefficients, variance-covariance parameters, and transition probabilities respectively.       

\begin{table}
\vspace{-3.5em}
  \centering
  \caption{Results of 500 simulations for experiments 1, 2 and 3 as described in the text.}
  \scriptsize
  \setlength{\tabcolsep}{14pt} 
  \renewcommand{\arraystretch}{0.8} 
  \begin{threeparttable}
  \begin{tabular}{rrrrrrrr}
    \toprule
                  & \multicolumn{3}{c}{Lasso}                     &               & \multicolumn{3}{c}{SCAD} \\
\cmidrule{2-4}\cmidrule{6-8}    \multicolumn{1}{l}{d} & \multicolumn{1}{c}{T = 100} & \multicolumn{1}{c}{T = 200} & \multicolumn{1}{c}{T = 300} &               & \multicolumn{1}{c}{T = 100} & \multicolumn{1}{c}{T = 200} & \multicolumn{1}{c}{T = 300} \\
    \midrule
    \multicolumn{1}{l}{Experiment 1} &               &               &               &               &               &               &  \\
    \multicolumn{1}{l}{\textit{True model included}} &               &               &               &               &               &               &  \\
    10            & 0.956         & 1.000         & 1.000         &               & 0.954         & 1.000         & 1.000 \\
    16            & 0.886         & 1.000         & 1.000         &               & 0.878         & 1.000         & 1.000 \\
    \multicolumn{1}{l}{\textit{Selected parameters}} &               &               &               &               &               &               &  \\
    10            & 70.50         & 65.72         & 56.87         &               & 71.70         & 59.17         & 55.29 \\
    16            & 167.62        & 123.53        & 101.54        &               & 146.25        & 107.03        & 90.55 \\
    \multicolumn{1}{l}{\textit{Share of non-zero}} &               &               &               &               &               &               &  \\
    10            & 0.998         & 1.000         & 1.000         &               & 0.997         & 1.000         & 1.000 \\
    16            & 0.996         & 1.000         & 1.000         &               & 0.996         & 1.000         & 1.000 \\
    \multicolumn{1}{l}{\textit{RMSE}} &               &               &               &               &               &               &  \\
    10            & 1.452         & 1.390         & 0.938         &               & 1.608         & 1.231         & 0.947 \\
    16            & 2.945         & 2.055         & 1.697         &               & 2.809         & 1.889         & 1.267 \\
    \multicolumn{1}{l}{\textit{$RMSE_{VAR}$}} &               &               &               &               &               &               &  \\
    10            & 1.227         & 0.813         & 0.683         &               & 0.949         & 0.488         & 0.313 \\
    16            & 1.728         & 1.069         & 0.914         &               & 1.346         & 0.648         & 0.482 \\
    \multicolumn{1}{l}{\textit{$RMSE_{COV}$}} &               &               &               &               &               &               &  \\
    10            & 0.776         & 1.128         & 0.643         &               & 1.298         & 1.130         & 0.894 \\
    16            & 2.384         & 1.755         & 1.430         &               & 2.465         & 1.774         & 1.172 \\
    \multicolumn{1}{l}{\textit{$RMSE_p$}} &               &               &               &               &               &               &  \\
    10            & 0.128         & 0.086         & 0.070         &               & 0.128         & 0.086         & 0.070 \\
    16            & 0.120         & 0.081         & 0.069         &               & 0.120         & 0.081         & 0.069 \\
    \midrule
    \multicolumn{1}{l}{Experiment 2} &               &               &               &               &               &               &  \\
    \multicolumn{1}{l}{\textit{True model included}} &               &               &               &               &               &               &  \\
    10            & 0.268         & 0.854         & 0.998         &               & 0.264         & 0.972         & 1.000 \\
    16            & 0.066         & 0.810         & 0.990         &               & 0.060         & 0.924         & 1.000 \\
    \multicolumn{1}{l}{\textit{Selected parameters}} &               &               &               &               &               &               &  \\
    10            & 98.89         & 89.55         & 90.87         &               & 95.58         & 91.53         & 102.63 \\
    16            & 188.84        & 157.81        & 138.47        &               & 181.29        & 167.15        & 165.00 \\
    \multicolumn{1}{l}{\textit{Share of non-zero}} &               &               &               &               &               &               &  \\
    10            & 0.947         & 0.998         & 1.000         &               & 0.945         & 0.999         & 1.000 \\
    16            & 0.930         & 0.996         & 1.000         &               & 0.928         & 0.999         & 1.000 \\
    \multicolumn{1}{l}{\textit{RMSE}} &               &               &               &               &               &               &  \\
    10            & 2.527         & 2.323         & 1.367         &               & 2.526         & 1.593         & 1.201 \\
    16            & 3.739         & 2.496         & 2.078         &               & 3.757         & 2.101         & 1.647 \\
    \multicolumn{1}{l}{\textit{$RMSE_{VAR}$}} &               &               &               &               &               &               &  \\
    10            & 1.517         & 0.942         & 0.666         &               & 1.533         & 0.808         & 0.465 \\
    16            & 2.110         & 1.350         & 1.057         &               & 2.116         & 1.148         & 0.733 \\
    \multicolumn{1}{l}{\textit{$RMSE_{COV}$}} &               &               &               &               &               &               &  \\
    10            & 2.021         & 2.124         & 1.194         &               & 2.007         & 1.374         & 1.107 \\
    16            & 3.087         & 2.100         & 1.789         &               & 3.104         & 1.760         & 1.475 \\
    \multicolumn{1}{l}{\textit{$RMSE_p$}} &               &               &               &               &               &               &  \\
    10            & 0.128         & 0.084         & 0.067         &               & 0.128         & 0.084         & 0.067 \\
    16            & 0.126         & 0.081         & 0.068         &               & 0.126         & 0.081         & 0.068 \\
    \midrule
    \multicolumn{1}{l}{Experiment 3} &               &               &               &               &               &               &  \\
    \multicolumn{1}{l}{\textit{True model included}} &               &               &               &               &               &               &  \\
    10            & 0.008         & 0.302         & 0.624         &               & 0.002         & 0.300         & 0.666 \\
    16            & 0.000         & 0.068         & 0.458         &               & 0.000         & 0.056         & 0.470 \\
    \multicolumn{1}{l}{\textit{Selected parameters}} &               &               &               &               &               &               &  \\
    10            & 273.39        & 256.94        & 238.20        &               & 264.71        & 252.06        & 233.77 \\
    16            & 604.87        & 587.33        & 546.63        &               & 565.69        & 569.72        & 534.27 \\
    \multicolumn{1}{l}{\textit{Share of non-zero}} &               &               &               &               &               &               &  \\
    10            & 0.937         & 0.984         & 0.994         &               & 0.918         & 0.984         & 0.995 \\
    16            & 0.906         & 0.978         & 0.994         &               & 0.848         & 0.975         & 0.994 \\
    \multicolumn{1}{l}{\textit{RMSE}} &               &               &               &               &               &               &  \\
    10            & 2.193         & 1.280         & 0.993         &               & 2.908         & 1.362         & 0.946 \\
    16            & 3.542         & 1.994         & 1.461         &               & 5.501         & 2.274         & 1.459 \\
    \multicolumn{1}{l}{\textit{$RMSE_{VAR}$}} &               &               &               &               &               &               &  \\
    10            & 1.901         & 1.149         & 0.908         &               & 2.646         & 1.220         & 0.845 \\
    16            & 2.982         & 1.740         & 1.315         &               & 5.037         & 2.012         & 1.289 \\
    \multicolumn{1}{l}{\textit{$RMSE_{COV}$}} &               &               &               &               &               &               &  \\
    10            & 1.092         & 0.563         & 0.400         &               & 1.207         & 0.604         & 0.426 \\
    16            & 1.912         & 0.973         & 0.636         &               & 2.210         & 1.060         & 0.683 \\
    \multicolumn{1}{l}{\textit{$RMSE_p$}} &               &               &               &               &               &               &  \\
    10            & 0.125         & 0.082         & 0.067         &               & 0.125         & 0.082         & 0.067 \\
    16            & 0.124         & 0.082         & 0.068         &               & 0.124         & 0.082         & 0.068 \\
    \bottomrule
    \end{tabular}%

    \begin{tablenotes}[flushleft]
    \small
    \item Notes: \textit{True model included} is the average share of estimates that includes all truly non-zero coefficients. \textit{Selected parameters} is the number of non-zero estimates. \textit{Share of non-zero} is the average share of truly non-zero coefficients that was estimated by the algorithm as non-zero. $RMSE = \sqrt{\frac{1}{200} \sum_{i=1}^{200} \| \hat{\phi}_T(i) - \phi_T^* \|}$ where $\hat{\phi}_T(i)$ is the estimated parameter vector for iteration $i$ containing all parameters in the system. $RMSE_{VAR}, RMSE_{COV},$ and $RMSE_{p}$ are similarly defined \textit{RMSE} measures for estimated VAR coefficients, variance-covariance parameters, and transition probabilities respectively.
    \end{tablenotes}
    \end{threeparttable}
  \label{tab:simulation}%
\end{table}%

The results of the experiments are presented in Table \ref{tab:simulation}. Looking at the metrics for estimation consistency, we see that all $RMSE$ measures are declining as the sample size increases for both Lasso and SCAD. It is interesting to note that SCAD appears to perform better for larger sample sizes in terms of estimation accuracy as measured by $RMSE$. Furthermore, we see that, in most cases, both estimators get better at including the true model while the number of selected parameters fall, which provides evidence of model selection consistency. We also note that SCAD has a slight advantage over the Lasso in selection accuracy for larger sample sizes as seen in experiments 2 and 3.

\section{Short-horizon stock return predictability}\label{returns}

It is well established that short-horizon stock return predictability (both in- and out-of-sample) exhibits significant time-variation \citep[e.g.][]{chen2012testing, rapach2013forecasting}. Specifically, \cite{henkel2011time} (henceforth HMN) and \cite{dangl2012predictive} show that return predictability is correlated with business cycles in a distinctively counter-cyclical fashion. In particular, HMN estimate a MS-VAR(1) with 2 states, one state corresponding to an 'expansion' regime and the other to a 'recession', for the period of 1953 to 2008 with excess returns, dividend yield, short rate, term spread, and default spread as endogenous variables for the US. With the estimated system, they calculate the adjusted $R^2$ ($\overline{R}^2$) from a predictive regression of lagged predictors on one-month ahead excess returns (the first equation from the VAR\footnote{Excess returns are ordered first in the system.}) for both model-implied periods of recession and expansion. The authors find that $\overline{R}^2$ is close to 0 during expansions while it averages around 0.175 for recessions. This led them to argue that return predictability is negligible during expansions and is present exclusively during recessions.

A major drawback of this strategy is that the predictors were selected in a relatively ad-hoc manner, which may yield significant omitted variable bias. The framework precludes the possibility that changes in return predictability are due to predictors beyond that of the chosen variables. This is a significant problem because if, as the authors argued, aggregate predictor variables are jointly determined by the "micro-level objectives of firms and central banks" which are in turn driven by business cycles, we should expect this counter-cyclical relationship between predictor and excess returns to potentially manifest in any predictor that relate to the "micromotives" of economic agents, which the literature on return predictability is not short of.

We approach this problem by incorporating 14 predictors as considered by \cite{welch2008comprehensive} in their study on return predictability: dividend price ratio (d/p), dividend yield (d/y), earnings price ratio (e/p), dividend payout ratio (d/e), stock variance (svar), book-to-market ratio (b/m), net equity expansion (ntis), treasury bill rate (tbl), long term rate of returns (ltr), long term yield (lty), term spread (tms), default yield spread (dfy), default return spread (dfr), and inflation (infl). Note that d/p, tbl, tms, and dfy overlap with the predictors in HMN. This results in a 2 state MS-VAR(1) system with 15 endogenous variables including excess returns (r) ordered first, with 711 estimable parameters including an intercept\footnote{To maintain the assumption of stationarity, we detrend all time series by subtracting a moving average of the past 12 months following \cite{ang2007stock}. Furthermore, as is customary in the literature on Lasso, we rescale all variables to have zero mean and a standard deviation of 1.}. Additionally, we update the investigation period to cover April 1953 to December 2018 (sample size of 787).  

\begin{figure}
\centering
\caption{Probability estimates of recession compared with an NBER-based recession indicator.}
\begin{threeparttable}
\includegraphics[scale=0.68, trim={0 1cm 0 0.5cm}, clip]{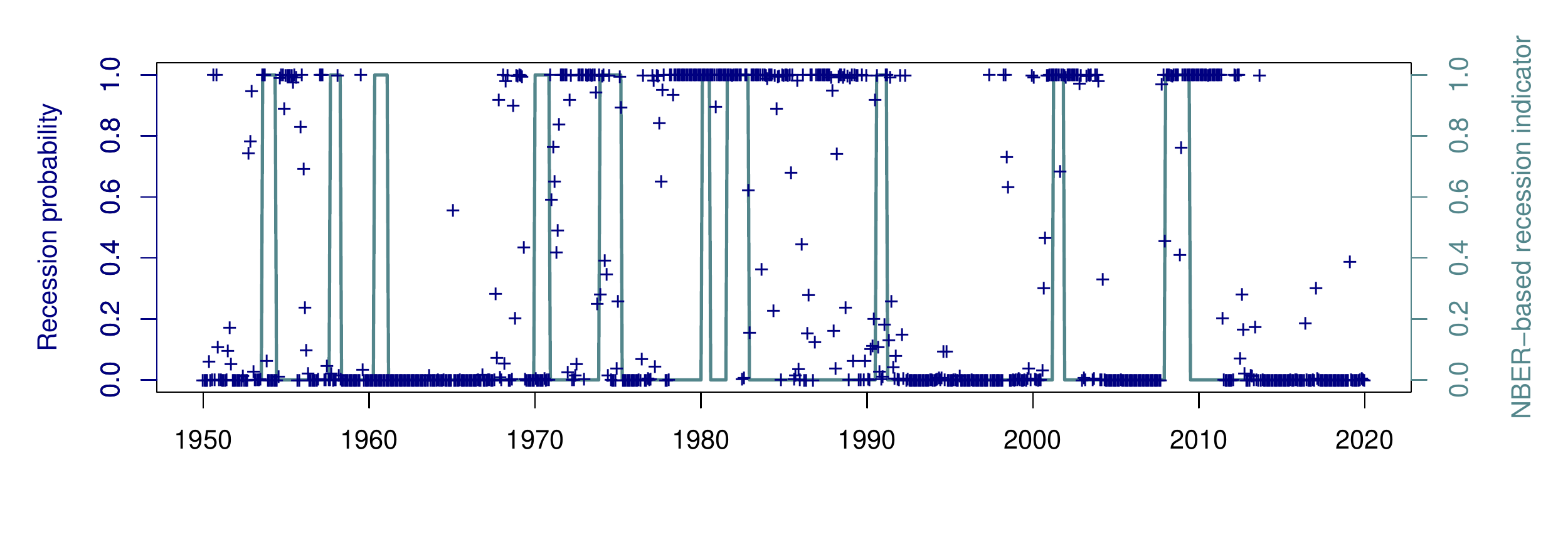}
\begin{tablenotes}[flushleft]
\small
\item Notes: (left axis, navy '+' labels) Smoothed probabilites from the MS-VAR for state 1 labelled by navy '+'; (right axis, turquoise line) Monthly NBER-based recession indicator from St. Louis FRED (USREC). A value of 1 indicates a recession, and 0 for expansion. 
\end{tablenotes}
\end{threeparttable}
\label{fig:rec_prob}
\end{figure}

Given the large number of parameters involved, the system is estimated with the SCAD penalty initialized with Lasso estimates. As a diagnostic check on the EM algorithm, we first verify that the estimated system yields regimes consistent with the interpretation of expansions and recessions. Figure \ref{fig:rec_prob} compares the smoothed probabilities for state 1 with that of an NBER-based recession indicator from the St. Louis FRED database (USREC). We see that when the probability of state 1 is close to unity, the NBER-based index often indicates a recession (value of 1), which provides evidence that state 1 corresponds to that of a recession. More formally, if we classify recessions as having a state 1 probability of greater than 0.5 and subsequently compare the resultant series with the NBER-based indicator, we get an agreement rate of 70\%, which is comparable to the 77\% obtained in HMN.

Several other features of our system are similar. The estimated transition probability matrix yields $p_{1\rightarrow 1} = 0.79$ (recession to recession) and $p_{2 \rightarrow 2} = 0.88$ (expansion to expansion), which are close to the respective estimates of $0.80$ and $0.91$ in HMN. We also replicate the finding of increased volatility for all predictors during recession relative to periods of expansion as reported in table \ref{tab:vol}.

\begin{table}
  \centering
  \caption{Differences in the volatility of predictors between periods of recessions and expansions.}
  \begin{threeparttable}
    \begin{tabular}{p{2cm}p{3cm}p{2cm}p{3cm}}
    \toprule
    \multicolumn{4}{l}{Difference in volatility: } \\
    \multicolumn{4}{l}{(estimated variance in state 1 - estimated variance in state 2)} \\
    \midrule
                  &               &               &  \\
    \multicolumn{1}{l}{r} & \multicolumn{1}{l}{0.844} & \multicolumn{1}{l}{ltr} & \multicolumn{1}{l}{0.952} \\
    \multicolumn{1}{l}{d/p} & \multicolumn{1}{l}{0.191} & \multicolumn{1}{l}{tms} & \multicolumn{1}{l}{0.410} \\
    \multicolumn{1}{l}{d/y} & \multicolumn{1}{l}{0.003} & \multicolumn{1}{l}{dfy} & \multicolumn{1}{l}{0.230} \\
    \multicolumn{1}{l}{e/p} & \multicolumn{1}{l}{0.125} & \multicolumn{1}{l}{dfr} & \multicolumn{1}{l}{1.101} \\
    \multicolumn{1}{l}{svar} & \multicolumn{1}{l}{1.687} & \multicolumn{1}{l}{infl} & \multicolumn{1}{l}{0.267} \\
    \multicolumn{1}{l}{b/m} & \multicolumn{1}{l}{0.286} & \multicolumn{1}{l}{d/e} & \multicolumn{1}{l}{0.088} \\
    \multicolumn{1}{l}{ntis} & \multicolumn{1}{l}{0.169} & \multicolumn{1}{l}{lty} & \multicolumn{1}{l}{0.388} \\
    \multicolumn{1}{l}{tbl} & \multicolumn{1}{l}{0.315} &               &  \\
                  &               &               &  \\
    \bottomrule
    \end{tabular}%
    \begin{tablenotes}[flushleft]
    \small
    \item Notes: As discussed in the main text, state 1 corresponds to recession while state 2 corresponds to expansions. r: excess return; d/p: dividend price ratio; d/y: dividend yield; e/p: earnings price ratio; svar: stock variance; b/m: book-to-market ratio; ntis: net equity expansion; tbl: treasury bill rate; ltr: long term rate of return; tms: term spread; dfy: default yield spread; dfr: default return spread; infl: inflation; d/e: dividend payout ratio; lty: long term yield. See \cite{welch2008comprehensive} for a detailed description of the variables.
    \end{tablenotes}
    \end{threeparttable}
  \label{tab:vol}%
\end{table}%

Panel A of Table \ref{tab:r2} reports the $\overline{R}^2$ derived from using the first equation of the MS-VAR (since $r$ is ordered first) in predicting one-month ahead excess returns conditional on state. These results present strong evidence in favor of the counter-cyclicality of return predictability given the larger magnitude of $\overline{R}^2$ during recessions (0.126) compared to expansion (0.112), which is in line with HMN. However, our estimate of $\overline{R}^2$ during expansions is at least 4 times larger than theirs, which is consistent with the findings of \cite{dangl2012predictive} that return predictability might still be present during booms, but are indeed stronger during busts.  

\begin{table}
  \centering
  \caption{Estimated $\overline{R}^2$, coefficients from predictive regressions of one-month ahead excess returns on lagged predictor variables for periods of expansion and recession, and out-of-sample forecasting results.}
  \footnotesize
  \setlength{\tabcolsep}{4pt} 
  \renewcommand{\arraystretch}{0.85}
  \begin{threeparttable}
             \begin{tabular}{rrrrrr}
    \toprule
    \multicolumn{1}{l}{Panel A:} & \multicolumn{5}{l}{Adjusted R2 from predictive regressions} \\
\cmidrule{2-6}          & \multicolumn{1}{l}{Expansion} &       & \multicolumn{1}{l}{Recession} &       &  \\
    \midrule
          & \multicolumn{1}{l}{0.112} &       & \multicolumn{1}{l}{0.126} &       &  \\
          &       &       &       &       &  \\
    \midrule
    \multicolumn{1}{l}{Panel B:} & \multicolumn{5}{l}{Coefficients of selected variables} \\
\cmidrule{2-6}          & \multicolumn{1}{l}{Expansion} &       & \multicolumn{1}{l}{Recession} &       &  \\
    \midrule
    \multicolumn{1}{l}{d/p} & \multicolumn{1}{l}{0.247} &       & \multicolumn{1}{l}{0.399} &       &  \\
    \multicolumn{1}{l}{svar} & \multicolumn{1}{l}{-0.040} &       & \multicolumn{1}{l}{-0.258} &       &  \\
    \multicolumn{1}{l}{tbl} & \multicolumn{1}{l}{-0.344} &       &       &       &  \\
    \multicolumn{1}{l}{e/p} &       &       & \multicolumn{1}{l}{0.019} &       &  \\
    \multicolumn{1}{l}{ltr} &       &       & \multicolumn{1}{l}{0.142} &       &  \\
    \multicolumn{1}{l}{tms} &       &       & \multicolumn{1}{l}{0.201} &       &  \\
    \multicolumn{1}{l}{dfr} &       &       & \multicolumn{1}{l}{0.062} &       &  \\
    \multicolumn{1}{l}{infl} &       &       & \multicolumn{1}{l}{-0.019} &       &  \\
    \multicolumn{1}{l}{lty} &       &       & \multicolumn{1}{l}{-0.245} &       &  \\
          &       &       &       &       &  \\
    \midrule
    \multicolumn{1}{l}{Panel C:} & \multicolumn{1}{l}{OOS Forecast Accuracy} &       & \multicolumn{3}{l}{Forecast comparison tests} \\
\cmidrule{2-2}\cmidrule{4-6}    \multicolumn{1}{l}{\textit{01.2010 - 12.2018}} & \multicolumn{1}{l}{MSFE} &       &       & \multicolumn{1}{l}{DM test} & \multicolumn{1}{l}{RC test} \\
    \midrule
    \multicolumn{1}{l}{MS-VAR (SCAD)} & \multicolumn{1}{l}{0.527} &       & \multicolumn{1}{l}{MS-VAR} &       &  \\
    \multicolumn{1}{l}{MS-VAR (HMN)} & \multicolumn{1}{l}{0.560} &       & \multicolumn{1}{l}{vs MS-VAR (HMN)} & \multicolumn{1}{l}{0.143} & \multicolumn{1}{l}{0.206} \\
    \multicolumn{1}{l}{Hist. Avg.} & \multicolumn{1}{l}{0.640} &       & \multicolumn{1}{l}{vs Hist. Avg.} & \multicolumn{1}{l}{0.044} & \multicolumn{1}{l}{0.074} \\
    \multicolumn{1}{l}{ARMA(1,1)} & \multicolumn{1}{l}{0.665} &       & \multicolumn{1}{l}{vs ARMA} & \multicolumn{1}{l}{0.033} & \multicolumn{1}{l}{0.057} \\
          &       &       &       &       &  \\
    \midrule
    \multicolumn{1}{l}{Panel D:} & \multicolumn{1}{l}{OOS Forecast Accuracy} &       & \multicolumn{3}{l}{Forecast comparison tests} \\
\cmidrule{2-2}\cmidrule{4-6}    \multicolumn{1}{l}{\textit{01.2007 - 12.2018}} & \multicolumn{1}{l}{MSFE} &       &       & \multicolumn{1}{l}{DM test} & \multicolumn{1}{l}{RC test} \\
    \midrule
    \multicolumn{1}{l}{MS-VAR (SCAD)} & \multicolumn{1}{l}{0.913} &       & \multicolumn{1}{l}{MS-VAR} &       &  \\
    \multicolumn{1}{l}{MS-VAR (HMN)} & \multicolumn{1}{l}{0.964} &       & \multicolumn{1}{l}{vs MS-VAR (HMN)} & \multicolumn{1}{l}{0.299} & \multicolumn{1}{l}{0.253} \\
    \multicolumn{1}{l}{Hist. Avg.} & \multicolumn{1}{l}{0.928} &       & \multicolumn{1}{l}{vs Hist. Avg.} & \multicolumn{1}{l}{0.468} & \multicolumn{1}{l}{0.480} \\
    \multicolumn{1}{l}{ARMA(1,1)} & \multicolumn{1}{l}{0.920} &       & \multicolumn{1}{l}{vs ARMA} & \multicolumn{1}{l}{0.482} & \multicolumn{1}{l}{0.465} \\
          &       &       &       &       &  \\
    \bottomrule
    \end{tabular}%
    \begin{tablenotes}[flushleft]
    \small
    \item Notes: Panel A: adjusted $R^2$ from the estimated predictive regression (first equation in MS-VAR) of one-month ahead excess returns on relevant lagged predictors as identified by the algorithm. Dates of expansion and recession are determined by estimated smooth probabilities of states.Panel B: Estimated coefficients from the first equation in the MS-VAR across the 2 states. Predictors that are not included here are regarded as not relevant and dropped by the algorithm. d/p: dividend price ratio; e/p: earnings price ratio; svar: stock variance; tbl: treasury bill rate; ltr: long term rate of return; dfy: default yield spread; dfr: default return spread; infl: inflation; lty: long term yield. See \cite{welch2008comprehensive} for a detailed description of the variables. Panel C and D: MSFE results for OOS period of Jan 2010 to Dec 2018 and Jan 2007 to Dec 2018 respectively. p-values of forecast comparison tests for Diebold and Mariano test (DM) \citep{diebold1995comparing} and 'Reality Check' test (RC) \citep{white2000reality} are included. "$x < y$" refers to a test of null hypothesis of equal predictive ability between $x$ and $y$, with the one-sided alternative of superior predictive ability of $x$ over $y$.
    \end{tablenotes}
    \end{threeparttable}
  \label{tab:r2}%
\end{table}%

A key merit of our regularized MS-VAR in this context is the ability to identify predictors that contribute the most to this counter-cyclical behavior of return predictability. Looking at Panel B of Table \ref{tab:r2}, we see that 2 predictors are relevant to both states: d/p, and svar. More importantly, we see that tbl drops out during times of crisis, which is reasonable given that the policy rate has recently been set constantly close to zero during protracted periods of economic slowdown. The inclusion of 6 new relevant predictors during recessions (e/p, ltr, tms, dfr, infl and lty) indicate that they contribute to return predictability during bad times only, which is a finding that is challenging to replicate in small MS-VARs with few endogenous variables. Interestingly, of all the selected variables, only d/p, tbl and tms were included in the MS-VAR of HMN, which suggests a potential omitted variable bias problem. Capturing these dynamics matter because they may help to consolidate the narrative on return predictability. For example, the non-relevance of e/p during expansions and its relevance during recessions is consistent with the idea of "accounting conservatism", as argued by HMN, which meant that balance sheets may be quicker in reporting bad news while reacting slower during good times. Slower disclosure and propagation of reported earnings information during expansions may hinder the predictive power of earnings ratios during such a period, while a faster response with bad news during crisis may lead to more accurate reporting of earnings, and contribute to predictive power.

Next, we study the out-of-sample (OOS) forecasting properties of the proposed large MS-VAR with an OOS period of January 2010 to December 2018. The model is estimated with an expanding window and one-step ahead forecasts are constructed as the weighted average of predictions from both states. To compare our results, we consider also the original 5 variable MS-VAR estimated in HMN. In addition, we form forecasts of $r$ using an expanding window historical average and include predictions from an ARMA(1,1) model. Specifically, the historical average is often hailed as notoriously difficult to beat \citep{welch2008comprehensive}. To formally compare the results, we employ two forecast comparison tests: Diebold and Mariano test \cite{diebold1995comparing} and the 'Reality check' test \citep{white2000reality}.

Panel C of Table \ref{tab:r2} reports the results. Notably, the proposed MS-VAR achieves the lowest mean squared forecast errors (MSFE) followed by the MS-VAR of HMN. Although the difference between the proposed MS-VAR and that of HMN is not statistically significant, we note that the large MS-VAR performs significantly better than both the historical average and ARMA(1,1), as indicated by both tests. 

Additionally, we consider an out-of-sample period encompassing the 2008 financial crisis in Panel D. Here, we expect regime-switching methods to perform better given that the possibility of having a recession state is explicitly accounted for. Indeed, we see that, relative to the historical average benchmark, this is the case for the SCAD-penalized MS-VAR but not for the original MS-VAR in HMN. One conjecture for the difference is that, as suggested in Panel B, many of the predictors that are relevant for predictability during recessions are included in the large MS-VAR but not the HMN model. This provides some evidence that prediction performance can be improved by considering more predictors. This is particularly the case because the large MS-VAR continues to perform the best in terms of achieving the lowest MSFE although this difference in predictive accuracy does not appear to be statistically significant. 

\section{Concluding remarks}

In this paper, we have proposed two new shrinkage type estimators to handle parameter proliferation in sparse high-dimensional MS-VARs. Theoretically, we have shown that both the Lasso and SCAD estimators are estimation consistent, while the latter has the added benefit of selecting relevant variables with high probability. Consequently, the SCAD estimator exhibits the oracle property in that it is asymptotically equivalent to an estimator that assumes \textit{a priori} knowledge of the sparsity pattern in the system. Results from numerical experiments show that the proposed EM algorithm is able to handle large MS-VARs well and the finite sample performance of the estimators provides support for our theoretical results. The empirical investigation on the counter-cyclicality of return predictability highlights the flexibility of the proposed estimation in incorporating many endogenous predictors and the merit of allowing for variable selection in regime-switching applications. Furthermore, the significantly better OOS performance of our model suggests that sizeable improvements to stock return prediction can be attained with a larger pool of predictors. Notably, this suggests that our framework can be generalized to other applications where regime-switching is of interest but where high-dimensionality may be a limiting factor, such as those frequently encountered in monetary policy, asset allocation, and other macroeconomic or financial systems.   

\newpage
\bibliographystyle{elsarticle-harv}
\bibliography{third_year_paper}
\processdelayedfloats

\newpage
\appendix
\section{Proofs for section \ref{asymptotic}}

Before proving \thref{consistency}, we require some intermediate results. We assume the following results hold under the conditions of \thref{consistency}. The proofs for these technical results can be found in an online appendix \citep{onlineappendix}.

\begin{lemma}\thlabel{momentresult}
The log-likelihood function evaluated at the true parameter, $\log \mathcal{L}_T(\phi_T^*)$, satisfies
\begin{align}
&E[\nabla^{1} \log \mathcal{L}_T(\phi_T^*)] = \boldsymbol{0} \label{score}\\
&E[(\nabla^{1} \log \mathcal{L}_T(\phi_T^*))(\nabla^{1} \log \mathcal{L}_T(\phi_T^*))^\top] = - E[\nabla^{2}  \log \mathcal{L}_T(\phi_T^*))]. \label{infomat}
\end{align}
\end{lemma}

Lemma 1 states that the Fisher score vector has zero mean when evaluated at the true parameter, and that the information matrix equality holds.

\begin{lemma}\thlabel{boundedmoments}
Let $\|\cdot\|_2$ represent the $L_2$ norm. We have $ \| \nabla^{1}_j \mathcal{L}_T(\phi_T^*)\|_2 \leq C_1$, $ \| \nabla^{2}_{j,k} \mathcal{L}_T(\phi_T^*)\|_2 \\ \leq  C_2$, and $ \| \nabla^{3}_{j,k,l} \mathcal{L}_T(\phi_T)\|_2 \leq C_3$ for $\phi_T \in \tilde{\Theta}_T$, where $C_i < \infty$ ($i=1,2,3$).
\end{lemma}

\begin{corollary}\thlabel{Oprate} 
Define $\partial^{1}_{o} \nabla^1_j \log \mathcal{L}(\phi_T) = \partial \nabla^1_j \log \mathcal{L}(\phi_T)/\partial \phi_T^{(\neg 0)}$, where $\phi_T^{(\neg 0)}$ refers to the parameters that are non-zero, and recall that $\nabla^1_j \log \mathcal{L}(\phi_T)$ is the $j^{th}$ element in $\nabla^{1} \log \mathcal{L}_T(\phi_T)$. Note that $\partial^{1}_{o} \nabla^1_j \log \mathcal{L}(\phi_T)$ is a $K_T^* \times 1$ vector. In addition, define $\partial^{2}_{o} \nabla^1_j \log \mathcal{L}(\phi_T) = \partial^2 \nabla^1_j \log \mathcal{L}(\phi_T)/\partial \phi_T^{{(\neg 0)}2}$ which is a $K_T^* \times K_T^*$ matrix. Then, we have the following probability bounds  
\begin{enumerate}
\item[(1)] $\| \nabla^{1} \log \mathcal{L}_T(\phi_T^*) \| = O_p(\sqrt{T K_T^*}),$
\item[(2)] $\| \nabla^{2} \log \mathcal{L}_T(\phi_T^*) - E[\nabla^{2}\log \mathcal{L}_T(\phi_T^*)] \|_{\textbf{1}} = O_p(\sqrt{T}K_T^*),$
\item[(3)] $\| \partial^{1}_{o} \nabla^1_j \log \mathcal{L}(\phi_T^*) - E[\partial^{1}_{o} \nabla^1_j \log \mathcal{L}(\phi_T^*)]\| = O_p(\sqrt{T K_T^*})$,
\item[(4)]  $ \| \nabla^{3} \log \mathcal{L}_T(\phi_T^*) \|_\textbf{1} = O_p(\sqrt{T} K_T^{*3/2}),$
\item[(5)] $\| \partial^{2}_{o} \nabla^1_j \log \mathcal{L}(\phi_T) \|_\textbf{1} = O_p(\sqrt{T} K_T^*)$ for $\phi_T \in \tilde{\Theta}_T$,
\end{enumerate}
where $\| A \|_{\textbf{1}}$ refers to the sum of all absolute values of elements.
\end{corollary}

\subsection{Proof of \thref{fsconst}}

First, following \cite{bickel2009simultaneous}, we bind the Lasso error with a cone constraint. For notation, equate the expression in \eqref{fstage} with $\mathcal{L}_T(\phi_T)^{FS}$. Then, by construction $\mathcal{L}_T(\tilde{\phi}_T)^{FS} \geq \mathcal{L}_T(\phi_T^*)^{FS}$. Hence,
\begin{align}
&T^{-1} [\log \mathcal{L}_T(\tilde{\phi}_T) -  \log \mathcal{L}_T(\phi_T^*)] \notag \\ 
&\geq \sum_{s=1}^{M}   \lambda^{Lasso}\bigg\{\sum_{m=1}^{d_T} \sum_{n=1}^{d_Tp_T} (|\tilde{a}_{mn}(s)| - |a^*_{mn}(s)|) + \sum_{m=1}^{d_T^*} \sum_{n=1}^{d_T^*q_T}(|\tilde{b}_{mn}(s)| - |b^*_{mn}(s)|)\bigg\} \label{VARcoef} \\ 
&+ \sum_{s=1}^M \lambda^{gLasso} \bigg\{ \sum_{m \neq n} (|\tilde{q}_{mn}(s)|-|q^*_{mn}(s)|)  \bigg\} \label{variance}
\end{align}
Let $S(h_T^A)$ and $S(h_T^B)$ be the set of indices where the VAR coefficients are non-zero across all states (see proof of \thref{consistency} below). So, we can rewrite \eqref{VARcoef} as
\begin{align}\label{a9}
&\lambda^{Lasso} \bigg\{ \bigg[\sum_{i \in S(h_T^A)} (|\tilde{a}_i| - |a^*_i|) + \sum_{i \notin S(h_T^A)} |\tilde{a}_i| \bigg] + \bigg[\sum_{i \in S(h_T^B)} (|\tilde{b}_i| - |b^*_i|) + \sum_{i \notin S(h_T^B)} |\tilde{b}_i| \bigg] \bigg\} \notag \\ 
&\geq \lambda^{Lasso} \bigg\{ \bigg[ -\sum_{i \in S(h_T^A)} |\Delta a_i| +  \sum_{i \notin S(h_T^A)} |\tilde{a}_i| \bigg] + \bigg[ -\sum_{i \in S(h_T^B)} |\Delta b_i| +  \sum_{i \notin S(h_T^B)} |\tilde{b}_i| \bigg] \bigg\}
\end{align} 
where we have defined $\Delta x_i \equiv \tilde{x}_i - x_i$. We can do the same for $S(h_T^{Qnd})$ and \eqref{variance}, to get
\begin{equation}\label{a10}
\eqref{variance} \geq \lambda^{gLasso} \bigg\{ -\sum_{i \in S(h_T^{Qnd})} |\Delta q_i| + \sum_{i \notin S(h_T^{Qnd})} |\tilde{q}_i| \bigg\} .
\end{equation}
On the other hand, by the local concavity assumption in \ref{itm:concave}, we get
\begin{equation*}
T^{-1} [\log \mathcal{L}_T(\tilde{\phi}_T) -  \log \mathcal{L}_T(\phi_T^*)] \leq \boldsymbol{\delta}^\top [T^{-1} \nabla^1 \log \mathcal{L}_T(\phi_T^*)], 
\end{equation*} 
where $\boldsymbol{\delta}$ is a vector of differences of parameter values at $\tilde{\phi}_T$ and $\phi^*_T$, which includes $((\Delta a_i)_i)$, $(\Delta b_i)_i$, and $(\Delta q_i)_i$. 

Note that from \thref{Oprate}, $T^{-1} \nabla^1 \log \mathcal{L}_T(\phi_T^*) = O_p(\sqrt{K_T^*/T}) = o_p(1)$, where the last equality comes from \ref{itm:rates}. Furthermore, observing that $\lambda^{Lasso} \propto \lambda^{gLasso}$, $\sum_{i \notin S(h_T^X)} |\tilde{x}_i| = \sum_{i \notin S(h_T^X)} |\Delta x_i|$, we conclude that with probability approaching one,
\begin{equation}\label{cone}
\sum_{i \notin S(h_T^A)} |\Delta a_i| + \sum_{i \notin S(h_T^B)} |\Delta b_i| + \sum_{i \notin S(h_T^{Qnd})} |\Delta q_i| \leq C \bigg\{ \sum_{i \in S(h_T^A)} |\Delta a_i| + \sum_{i \in S(h_T^B)} |\Delta b_i| +  \sum_{i \in S(h_T^{Qnd})} |\Delta q_i| \bigg\},
\end{equation}
where $C > 1$ is some constant.

\cite{bickel2009simultaneous} show that this cone constraint on the Lasso error in \eqref{cone} is necessary for proving convergence of the Lasso estimator. 

Now, we show \eqref{lassobound}. Define $\gamma_T^{FS} = \sqrt{K_T^*}\lambda^{Lasso}$ and recall the definition of $\Omega(W)$ from \ref{itm:concave}.  We want to show that there exists a local maximizer in the set $\tilde{\Theta}_{T,\gamma^{FS},W}$. To begin, note that we can write $\sum_{s=1}^M \sum_{m=1}^{d_T} \sum_{n=1}^{d_T p_T} |a_{mn}(s)| = \sum_i a_i$, where $a_i$ are elements of $\theta_A$. We can do the same for $b_{mn}(s)$ and $q_{mn}(s)$. For simplicity, let $\#A$ denote the dimensions of $A$. Then, pick any $\boldsymbol{u} \in \Omega(W)$ and observe that
\begin{align*}
\mathcal{L}_T(\phi_T^* + \gamma_T^{FS} \boldsymbol{u})^{FS} - \mathcal{L}_T(\phi_T^*)^{FS} &= T^{-1} [\log \mathcal{L}_T(\phi_T^* + \gamma_T^{FS} \boldsymbol{u}) - \log \mathcal{L}_T(\phi_T^*)] \\
&- \lambda^{Lasso}\bigg\{\sum_{i=1}^{\#\theta_A}(|a_i^* + \gamma_T^{FS}\boldsymbol{u}_{1i}| - |a_i^*|) + \sum_{i=1}^{\#\theta_B} (|b_i^* + \gamma_T^{FS}\boldsymbol{u}_{2i}| - |b_i^*|)\bigg\} \\
&- \lambda^{gLasso} \bigg\{ \sum_{i=1}^{\#Q^{ND}}(|q_i^* + \gamma_T^{FS}\boldsymbol{u}_{3i}| - |q_i^*|) \bigg\} \\
&\equiv I_1 + I_2 + I_3.
\end{align*}
For $I_1$, by the mean value theorem, for a $\check{\phi}_T$ between $\phi_T^* + \gamma_T^{FS}\boldsymbol{u}$ and $\phi_T^*$,
\begin{equation*}
I_1 = \gamma_T^{FS}\boldsymbol{u}^\top [T^{-1} \nabla^{1} \log \mathcal{L}_T(\phi_T^*)] + \frac{1}{2} \gamma_T^{FS^{2}}\boldsymbol{u}^\top [T^{-1}  \nabla^{2} \log \mathcal{L}_T(\check{\phi}_T)] \boldsymbol{u} \equiv I_{1,1} + I_{1,2}
\end{equation*}
Then,
\begin{equation*}
I_{1,1} \leq \gamma_T^{FS}\|\boldsymbol{u}\|O_p(\sqrt{K_T^*/T}) = \gamma_T^{FS}\|\boldsymbol{u}\| o_p(\sqrt{K_T^*}\lambda^{Lasso}) = \|\boldsymbol{u}\| o_p(\gamma_T^{FS^{2}}) 
\end{equation*}
where we have used \thref{Oprate} and \ref{itm:rates}. Next,
\begin{equation*}
I_{1,2} = \frac{1}{2} \gamma_T^{{FS}^2} \boldsymbol{u}^\top [T^{-1}\{\nabla^2 \log \mathcal{L}_T(\check{\phi}_T) - \nabla^2 \log \mathcal{L}_T(\phi_T^*)\}]\boldsymbol{u} + \frac{1}{2} \gamma_T^{{FS}^2} \boldsymbol{u}^\top \nabla^2 \log \mathcal{L}_T(\phi_T^*)\boldsymbol{u}.
\end{equation*}
Using the concavity assumption in \ref{itm:third}, we have
\begin{align*}
T^{-1} \gamma_T^{{FS}^2} \boldsymbol{u}^\top [\{\nabla^2 \log \mathcal{L}_T(\check{\phi}_T) - \nabla^2 \log \mathcal{L}_T(\phi_T^*)\}]\boldsymbol{u} &\leq \gamma_T^{{FS}^2}\nabla^{1}(\boldsymbol{u}^\top [T^{-1}  \nabla^{2} \log \mathcal{L}_T(\phi^*_T)] \boldsymbol{u}) \|\check{\phi}_T -  \phi_T^*\| \\
&\leq O_p(K_T^{*3/2}/\sqrt{T}) \|\boldsymbol{u}\|^3 \gamma_T^{{FS}^3} \\
&= o_p(\gamma_T^{{FS}^2})\|\boldsymbol{u}\|^2,
\end{align*}
where we have used (4) in \thref{boundedmoments} for the penultimate inequality. Next, observe that by the restricted eigenvalue condition in \ref{itm:eigen}, we get
\begin{equation*}
 \frac{1}{2} \gamma_T^{{FS}^2} \boldsymbol{u}^\top \nabla^2 \log \mathcal{L}_T(\phi_T^*)\boldsymbol{u} = - \frac{1}{2} \gamma_T^{{FS}^2} \boldsymbol{u}^\top [- \nabla^2 \log \mathcal{L}_T(\phi_T^*)]\boldsymbol{u} \leq - \frac{1}{2} \gamma_T^{{FS}^2} \rho_1 \|\boldsymbol{u}\|^2 < 0. 
\end{equation*}
So we conclude that the leading term in $I_{1,2}$ is negative.

Next, recalling that $\lambda^{gLasso} \propto \lambda^{Lasso}$, we have
\begin{align*}
I_3 &\leq \lambda^{gLasso} \sum_{i=1}^{h_T^{Qnd}} (|q_i^* + \gamma_T^{FS}\boldsymbol{u}_{3i}| - |q_i^*|) -  \lambda^{gLasso} \sum_{i=h_T^{Qnd}+1}^{\#Q^{ND}} |\gamma_T^{FS}\boldsymbol{u}_{3i}| \\
&= O_p(\gamma_T^{{FS}^2})\|\boldsymbol{u}_{3}\| - \lambda^{gLasso} \sum_{i=h_T^{Qnd}+1}^{\#Q^{ND}} |\gamma_T^{FS}\boldsymbol{u}_{3i}|,
\end{align*}
where the final leading term is negative. Furthermore, the same procedure can be applied to $I_2$ to yield similar results.

Hence the leading terms in $I_{1,2}$, $I_2$, and $I_3$ imply that for any $\varepsilon >0$, we can find a large $W >0$, such that 
\begin{equation*}
P\big(\sup_{\boldsymbol{u} \in \Omega(W)} \mathcal{L}_T(\phi_T^* + \gamma_T^{FS} \boldsymbol{u})^{FS} < \mathcal{L}_T(\phi_T^*)^{FS}\big) \geq 1-\varepsilon.
\end{equation*}
Therefore, we can find a local maximizer in $\tilde{\Theta}_{T,\gamma^{FS},W}$. Given the assumption of local concavity over $\tilde{\Theta}_{T} \supseteq \tilde{\Theta}_{T,\gamma^{FS},W}$, and that $\tilde{\phi}_T, \phi_T^* \in \tilde{\Theta}_{T}$, we obtain the result in \eqref{lassobound}.  
\qed

\subsection{Proof of \thref{consistency}}

Before approaching the proof proper, we define the following oracle problem, that is the maximum likelihood estimation of the model assuming \textit{a priori} knowledge of the sparsity pattern. Since this is unknowable in practice, this discussion is theoretical and functions as a device to judge the accuracy of our variable selection procedure. 

Let $\phi_T^{o}$ represent the vector where the irrelevant VAR parameters have been fixed at 0 prior to estimation. We define $S(h)$, for $h = h_T^A, h_T^B$ and $h_T^{Qnd}$, to be the set of indices whose corresponding parameter is non-zero, and $a_i$ is the $i^{th}$ element of $\theta_A$, likewise for $b_i$ ($\theta_B$) and $q_i$ ($Q^{ND}$). The (penalized) oracle problem is
\begin{equation}\label{oracleproblem}
\max_{\phi^{o}_T} T^{-1} \log \mathcal{L}(\mathcal{Y}_T| \mathcal{X}_T; \phi^{o}) - \bigg\{ \sum_{i \in S(h_T^A)} p^{'}_{\lambda}(|\tilde{a}_i|)|a_i| +  \sum_{i \in S(h_T^B)} p^{'}_{\lambda}(|\tilde{b}_i|)|b_i| + \sum_{i \in S(h_T^{Qnd})} p^{'}_{\lambda}(|\tilde{q}_i|)|q_i| \bigg\}.
\end{equation}
Notice that this is essentially the same problem as \eqref{sstage} but with irrelevant parameters set to 0 before estimation.

Suppose $\hat{\phi}_T^{o}$ is the unique solution to \eqref{oracleproblem}. To ensure that the dimensions of $\hat{\phi}_T^{o}$ and $\phi_T^{*}$ match, we define
\begin{equation}\label{oracleest}
\bar{\phi}_T^{o} = (\hat{\phi}_1, \ldots, \hat{\phi}_{K_T^{Sp}}, 0, \ldots, 0, \hat{q}_{11}(1), \ldots, \hat{q}_{d_Td_T}(M), \hat{p}_{1\shortrightarrow 1}, \ldots, \hat{p}_{M \shortrightarrow M})^\top,
\end{equation}
where parameters with hats are from $\hat{\phi}_T^{o}$.

\noindent \textbf{Proof of (1)}

Define $\gamma_T^{SS}=\sqrt{K_T^*/T}$ and construct the set $\tilde{\Theta}_{T,\gamma^{SS},W}$ for a sufficiently large constant $W$. Since $1/\sqrt{T}=o(\lambda^{Lasso})$, observe that $\tilde{\Theta}_{T,\gamma^{SS},W} \subseteq \tilde{\Theta}_{T,\gamma^{FS},W}$. Pick $\boldsymbol{u}^{*} \in \Omega(W)$, such that the elements $\boldsymbol{u}^{*}_{ji} = 0$ for $j=1,2,3$ and $i \notin S(h_T^X)$ where $X$ is either $\theta_A, \theta_b,$ or $Q^{ND}$. Define $\mathcal{L}_T(\phi_T)^{SS}$ to represent the expression \eqref{sstage} in the SCAD problem. Consider
\begin{align*}
\mathcal{L}_T(\phi_T^* + \gamma_T^{SS}\boldsymbol{u}^*)^{SS} &- \mathcal{L}_T(\phi_T^*)^{SS} = T^{-1} [\log \mathcal{L}_T(\phi_T^* + \gamma_T^{SS}\boldsymbol{u}^*) - \log \mathcal{L}_T(\phi_T^*)]   \\
&+ \bigg[ \sum_{i=1}^{\#\theta_A} p_\lambda^{'}(|\tilde{a}_i|) \bigg( |a_i^*| - |a_i^* + \gamma_T^{SS}\boldsymbol{u}^*_{1i}| \bigg) + \sum_{i=1}^{\#\theta_B} p_\lambda^{'}(|\tilde{b}_i|) \bigg( |b_i^*| - |b_i^* + \gamma_T^{SS}\boldsymbol{u}^*_{2i}| \bigg) \bigg] \\
&+ \sum_{i=1}^{\#Q^{ND}} p_{\lambda^*}^{'} (|\tilde{q}_i|) \bigg( |q_i^*| - |q_i^* + \gamma_T^{SS}\boldsymbol{u}^*_{3i}| \bigg) \\
&\equiv I_1^S + I_2^S + I_3^S. 
\end{align*}
Again, notice that since $1/\sqrt{T}=o(\lambda^{Lasso})$, the proof from \thref{fsconst} of $I_1$ is directly applicable to $I_1^S$ here and this shows that the leading terms are both, of an order larger than $\gamma_T^{{SS}^2}$, and are negative. 

Next, consider $I_3^S$. By construction of $\boldsymbol{u}^*$, we have
\begin{equation*}
I_3^S = \sum_{i=1}^{h_T^{Qnd}} p_{\lambda^*}^{'} (|\tilde{q}_i|) \bigg( |q_i^*| - |q_i^* + \gamma_T^{SS}\boldsymbol{u}^*_{3i}| \bigg),
\end{equation*}
since $q_i^*=0$ and $\boldsymbol{u}_{3i}^* = 0$ for $i > h_T^{Qnd}$. Next, note that 
\begin{equation}
|\tilde{q}_i| \geq \min_{1 \leq i \leq h_T^{Qnd}} |\tilde{q}_i| - \max_{1 \leq i \leq h_T^{Qnd}}  |\tilde{q}_i - q_i^*| \geq \lambda + o_p(\lambda), \label{SCADasym}
\end{equation}
where we have used \ref{itm:nonzero} for the first term, and \thref{fsconst} for the second term to show that it is $O_p(\sqrt{K_T^*}\lambda^{Lasso}) = o_p(\lambda)$, where the last equivalence follow from assumption \ref{itm:rates}. By the structure of the SCAD penalty in \eqref{SCADpenalty}, $p_\lambda^{'} = \theta$ if $0 < \theta \leq \lambda$ and $0$ if $\theta > \lambda$. Since $\lambda^* \propto \lambda$, we can say that $I_3^S = o_p(\gamma_T^{{SS}^2})\|\boldsymbol{u}_3^*\|$. Likewise, we have that $I_2^{S} = o_p(\gamma_T^{{SS}^2})(\|\boldsymbol{u}_1^*\|+\|\boldsymbol{u}_2^*\|)$.

Taken together, we arrive at the existence of a local maximizer in the ball $\tilde{\Theta}_{T,\gamma^{SS},W} \subseteq \tilde{\Theta}_T$, where the log likelihood is locally concave on $\tilde{\Theta}_T$.

\noindent \textbf{Proof of (2)}

Recall that $\bar{\phi}_T^{o}$ is defined in \eqref{oracleest}, which is the solution to a problem where we know the true sparsity pattern. Hence, we can complete the proof by showing $P(\hat{\phi}_T = \bar{\phi}_T^{o}) \rightarrow 1$.

For ease of exposition, we emphasize that the sub-vector $(\hat{\phi}_1,\ldots,\hat{\phi}_{K_T^{Sp}})$ is made up of the following vectors: $(\hat{a}_i)_i$, $(\hat{b}_i)_i$, and $(\hat{q}_i)_i$ for $i \in S(h_T^A),S(h_T^B)$, and $S(h_T^{Qnd})$ respectively, where $\hat{x}_i$ for $x=a,b,q$ refers to estimated parameters in $\theta_A$, $\theta_B$ and $Q^{ND}$. On the other hand, the subvector of zeros in $\bar{\phi}_T^{o}$ corresponds to $\hat{x}_i=0$ for $i \notin S(h_T^A),S(h_T^B)$, and $S(h_T^{Qnd})$.

To show that $\bar{\phi}_T^{o}$ is the solution to \eqref{sstage} with probability approaching one, we need to verify that it satisfies the following Karush-Kuhn-Tucker conditions:
\begin{enumerate}[label=(\Roman*)]
\item $T^{-1} \frac{\partial  \log \mathcal{L}_T(\phi_T)}{\partial a_i} \big|_{\hat{a}_i} - p_{\lambda}^{'}(|\tilde{a}_i|) \frac{\hat{a}_i}{|\hat{a}_i|}  = 0$, \inlineitem $T^{-1} \frac{\partial \log \mathcal{L}_T(\phi_T)}{\partial b_i} \big|_{\hat{b}_i} - p_{\lambda}^{'}(|\tilde{b}_i|) \frac{\hat{b}_i}{|\hat{b}_i|}  = 0$,

\item $T^{-1} \frac{\partial \log \mathcal{L}_T(\phi_T)}{\partial q_i} \big|_{\hat{q}_i} - p_{\lambda^*}^{'}(|\tilde{q}_i|) \frac{\hat{q}_i}{|\hat{q}_i|}  = 0$, \inlineitem $T^{-1} \frac{\partial \log \mathcal{L}_T(\phi_T)}{\partial q^{D}_i} \big|_{\hat{q}_i^{D}} = 0$,

\item $T^{-1} \frac{\partial \log \mathcal{L}_T(\phi_T)}{\partial \pi_i} \big|_{\hat{\pi}_i}  = 0$,
\end{enumerate}
where $q_i^D$ and $\pi_i$ are parameters from $Q^{D}$ and $\pi$. Note that by construction of $\bar{\phi}_T^{o}$, (IV) and (V) automatically holds for all $i$, and (I)-(III) holds for $i \in S(h_T^A),S(h_T^B)$, and $S(h_T^{Qnd})$ respectively. For indices $i$ that do not fall in those sets (i.e. the zeroes), we have to verify that the sub-differentials satisfy:
\begin{enumerate}[label=(\roman*)]
\item $\max_{i \notin  S(h_T^A)} \big| T^{-1} \frac{\partial  \log \mathcal{L}_T(\phi_T)}{\partial a_i} \big|_{\hat{a}_i} \leq \min_{i \notin  S(h_T^A)} p_{\lambda}^{'} (|\tilde{a}_i|)$, 

\item $\max_{i \notin  S(h_T^B)} \big| T^{-1}  \frac{\partial  \log \mathcal{L}_T(\phi_T)}{\partial b_i} \big|_{\hat{b}_i} \leq \min_{i \notin  S(h_T^B)} p_{\lambda}^{'} (|\tilde{b}_i|)$,

\item $\max_{i \notin  S(h_T^{Qnd})} \big| T^{-1}  \frac{\partial  \log \mathcal{L}_T(\phi_T)}{\partial q_i} \big|_{\hat{q}_i} \leq \min_{i \notin  S(h_T^{Qnd})} p_{\lambda^*}^{'} (|\tilde{q}_i|)$.
\end{enumerate}
We prove (i)-(iii) more generally by looking at $\max_{j \leq K_T} T^{-1}|\nabla^{1}_j \log \mathcal{L}_T(\hat{\phi}_T)|$ where $K_T$ is the total number of candidate parameters. Recall the definition of $\partial^{1}_{o} \nabla^1_j \log \mathcal{L}(\phi_T)$ and $\partial^{2}_{o} \nabla^1_j \log \mathcal{L}(\phi_T)$ from \thref{Oprate}. We want to show that $\max_{j \leq K_T} T^{-1}|\nabla^{1}_j \log \mathcal{L}_T (\hat{\phi}_T)| = o_p(\lambda)$, and we do so by adopting a similar approach to \cite{kwon2012large}.

Observe that
\begin{align*}
P(\max_{j \leq K_T} |\nabla^{1}_j \log &\mathcal{L}_T(\hat{\phi}_T)| > T\lambda) \leq P(\max_{j \leq K_T} |\nabla^{1}_j \log \mathcal{L}_T(\phi_T^*)| > T\lambda/4) \\ 
&+ P(\max_{j \leq K_T} \| \partial^{1}_{o} \nabla^1_j \log \mathcal{L}(\phi_T) - E[\partial^{1}_{o} \nabla^1_j \log \mathcal{L}(\phi_T)]\|\| \hat{\phi}_T - \phi_T^* \| > T\lambda/4) \\
&+ P(\max_{j \leq K_T}  \| E[\partial^{1}_{o} \nabla^1_j \log \mathcal{L}(\phi_T)]\|\| \hat{\phi}_T - \phi_T^* \| > T\lambda/4) \\
&+ P(\max_{j \leq K_T}  \|\partial^{2}_{o} \nabla^1_j \log \mathcal{L}(\phi_T)\|_\textbf{1} \| \hat{\phi}_T - \phi_T^* \|^2 > T\lambda/2 ) \\
&\equiv J_1 + J_2 + J_3 + J_4.
\end{align*}
To begin,
\begin{equation*}
J_1 \leq \sum_{i = K_T^*+1}^{K_T} P\bigg(|\nabla^{1}_j \log \mathcal{L}_T(\phi_T^*)| > T\lambda/4\bigg) = O\bigg(\frac{K_T}{T \lambda^2}\bigg) = o(1).
\end{equation*} 
Next,
\begin{align*}
J_2 &\leq P\bigg(\max_{j \leq K_T} \| \partial^{1}_{o} \nabla^1_j \log \mathcal{L}(\phi_T) - E[\partial^{1}_{o} \nabla^1_j \log \mathcal{L}(\phi_T)] \| > T\sqrt{T}\lambda/4 \sqrt{K_T^*} \bigg)  \\ 
&+ P\bigg(\| \hat{\phi}_T - \phi_T^* \| > \sqrt{K_T^*/T}\bigg) \\
&= O\bigg(\frac{K_T}{T^2 \lambda^2 / K_T^{*2}}\bigg) + o(1) = o(1),
\end{align*}
where we have use (3) from \thref{Oprate} for the first term, and the first part of \thref{consistency} for the second term.

For $J_3$, since $E[\partial^{1}_{o} \nabla^1_j \log \mathcal{L}(\phi_T)]$ is bounded as per \thref{boundedmoments}, we have
\begin{equation*}
J_3 \leq P\bigg(\|\hat{\phi}_T - \phi_T^* \| > T\lambda/4\sqrt{K_T^*}C\bigg) = o(1)
\end{equation*}
where the last equality is again by part (1) of \thref{consistency}.

Finally, using (5) in \thref{Oprate},
\begin{align*}
J_4 &\leq P\bigg(\max_{j \leq K_T}  \|\partial^{2}_{o} \nabla^1_j \log \mathcal{L}(\phi_T)\|_\textbf{1} > T^{*2}\lambda/2K_T^*\bigg) + P\bigg(\| \hat{\phi}_T - \phi_T^* \|^2 > K_T^*/T\bigg) \\
&= O\bigg(\frac{K_T}{T^2\lambda^2/K_T^{*4}} \bigg) + o(1) = o(1).
\end{align*}
Therefore, we have $\max_{j \leq K_T} T^{-1} |\nabla^{1}_j \log \mathcal{L}_T(\hat{\phi}_T)| = o_p(\lambda)$.

Next, recall that $|\tilde{\phi}_i|$ is $O_p(\sqrt{K_T^*}\lambda^{Lasso})$ by \thref{fsconst}. Furthermore, \ref{itm:rates} imply that $\sqrt{K_T^*}\lambda^{Lasso} = o(\lambda)$. Then, by the structure of the SCAD penalty where $p^{'}_\lambda(\theta) = \theta$ for $0 < \theta \leq \lambda$ or 0 otherwise, we have $\min_{j \leq K_T} p^{'}_\lambda (|\tilde{\phi}_j|) = \lambda$.

Taken together, this implies that the inequalities (i)-(iii) hold, which establishes part (2) of \thref{consistency}.

\qed 

\subsection{Proof of \thref{norm}}

First we consider the unpenalized oracle sub-problem given as
\begin{equation}
\max_{\phi^{o}_T} T^{-1} \log\mathcal{L}(\mathcal{Y}_T|\mathcal{X}_T; \phi^{o}). \label{oraclesubproblem}
\end{equation}
Note that this problem is identical to the oracle problem in \eqref{oracleproblem} but with the penalties removed. Let $\bar{\bar{\phi}}^{o}_T$ be the solution to \eqref{oraclesubproblem}. The proof can be accomplished by showing the following results: 
\begin{enumerate}
\item[(1)] 
\begin{equation*}
\|\hat{\phi_T}^{(\neg 0)} - \bar{\bar{\phi}}_T^{o}\| = o_p(T^{-1/2}). \label{oracleconst}
\end{equation*}

\item[(2)] 
\begin{equation*}\label{asymptoticnorm2}
\sqrt{T} G_T [I_T^{(\neg 0)}(\phi_T^*)]^{1/2}(\bar{\bar{\phi}}^{o}_T - \phi_T^{*(\neg0)}) \rightarrow^{d} \mathcal{N}(0,G).
\end{equation*}
\end{enumerate}

Intuitively, part (1) implies that $\hat{\phi}_T$ is asymptotically close to $\bar{\bar{\phi}}^{o}_T$ while part (2) shows that $\bar{\bar{\phi}}^{o}_T$ is asymptotically normal. Together, these results suggests that the same statistical inference for $\bar{\bar{\phi}}^{o}_T$ can be applied to $\hat{\phi}_T$, and that it shares the same efficiency as the oracle estimator.

\noindent \textbf{Proof for (1)}

With a slight abuse of notation, for this proof, we let $\phi^*_T = \phi_T^{*(\neg 0)}$.

In light of part (2) in \thref{consistency}, the penalized oracle estimator, $\bar{\phi}_T^{o}$ is indeed the solution to the SCAD problem initialized with Lasso estimates with probability approaching one. Hence, it suffices to check $\|\hat{\phi}_T^{o} - \bar{\bar{\phi}}_T^{o}\|=o_p(T^{-1/2})$, where recall that $\hat{\phi}_T^{o}$ is the solution to the penalized oracle problem in \eqref{oracleproblem} (i.e. $\bar{\phi}_T^{o}$ with the zeroes removed).  

Define $\dot{\lambda}=\big(p_\lambda^{'}(|\tilde{a}_1|)\frac{\hat{a}_1}{|\hat{a}_1|}, \ldots, p_\lambda^{'}(|\tilde{a}_{h_T^A}|)\frac{\hat{a}_{h_T^A}}{|\hat{a}_{h_T^A}|}, p_\lambda^{'}(|\tilde{b}_1|)\frac{\hat{b}_1}{|\hat{b}_1|}, \ldots,   p_{\lambda^{*}}^{'}(|\tilde{q}_{h_T^{Qnd}}|)\frac{\hat{q}_{h_T^{Qnd}}}{|\hat{q}_{h_T^{Qnd}}|}\big)^\top$. Then, note that for a $\check{\phi}_T^{b}$ between $\hat{\phi}_T^{o}$ and $\phi^*_T$, $\hat{\phi}_T^{o}$ satisfies
\begin{equation}
T^{-1} \nabla^{1} \log \mathcal{L}_T(\phi_T^*) = T^{-1} \nabla^2 \log \mathcal{L}_T(\check{\phi}_T^{b})(\phi_T^* - \hat{\phi}_T^{o}) - \dot{\lambda}. \label{FOCbiased}
\end{equation}
Next, for $\check{\phi}_T$ between $\bar{\bar{\phi}}_T^{o}$ and $\phi^*_T$,  $\bar{\bar{\phi}}_T^{o}$ satisfies
\begin{equation}
T^{-1} \nabla^{1} \log \mathcal{L}_T(\phi_T^{*}) = T^{-1} \nabla^{2} \log \mathcal{L}_T(\check{\phi}_T) (\phi_T^* - \bar{\bar{\phi}}_T^{o}). \label{FOCunbiased}
\end{equation}
Taking the difference of \eqref{FOCbiased} and \eqref{FOCunbiased}, we get
\begin{equation}
\dot{\lambda} = [T^{-1}(\nabla^2 \log \mathcal{L}_T (\check{\phi}_T^{b}) - \nabla^2 \log \mathcal{L}_T (\check{\phi}_T))] \phi_T^* + T^{-1} \nabla^2 \log \mathcal{L}_T (\check{\phi}_T) \bar{\bar{\phi}}_T^{o} - T^{-1} \nabla^2 \log \mathcal{L}_T (\check{\phi}_T^{b}) \hat{\phi}_T^{o}. \label{difference}
\end{equation}
Note that by \ref{itm:third} and (4) from \thref{Oprate},
\begin{align}
T^{-1}(\nabla^2 \log \mathcal{L}_T (\check{\phi}_T^{b}) - \nabla^2 \log \mathcal{L}_T (\check{\phi}_T))\phi^*_T &= [T^{-1}(\nabla^2 \log \mathcal{L}_T (\check{\phi}_T^{b}) - \nabla^2 \log \mathcal{L}_T (\phi_T^*))]\phi_T^* \notag \\
&+ [T^{-1}(\nabla^2 \log \mathcal{L}_T (\phi_T^*) - \nabla^2 \log \mathcal{L}_T (\check{\phi}_T))]\phi_T^* \notag \\
&= O_p(K_T^{*2}/T). \label{extraterm}
\end{align} 
Furthermore, By the same logic in \eqref{SCADasym}, the term $\dot{\lambda}$ goes to zero. Under the assumptions of \ref{itm:stat}-\ref{itm:minorization}, we can use Lemma 2 of \cite{bickel1998asymptotic} to get $T^{-1} \nabla^2 \log \mathcal{L}_T(\phi) \rightarrow^p - I_T^{\neg 0}(\phi_T^*)$ where $\phi$ refers to either $\check{\phi}_T^{b}$ or $\check{\phi}_T$ since both converge to $\phi^*_T$ in probability. Hence, $T^{-1} \nabla^2 \log \mathcal{L}_T(\phi) = O_p(1)$. Therefore, we get $\|\bar{\bar{\phi}}_T^{o} - \hat{\phi}_T^{o}\| = o_p(T^{-1/2})$ by noting that $\sqrt{T} O_p(K_T^{*2}/T) \\ = o_p(1)$ in \eqref{extraterm}, and using \eqref{difference}.

\noindent \textbf{Proof for (2)}

Here, we cannot directly apply the result or proof of asymptotic normality from low-dimensional papers such as \cite{bickel1998asymptotic} because of the diverging number of parameters. Instead, we approach it with the Lindeberg-Feller CLT as in \cite{fan2004nonconcave}. 

We start with \eqref{FOCunbiased}. Pre-multiply the left hand side with $\sqrt{T} G_T [I_T^{(\neg 0)}(\phi_T^*)]^{-1/2}$ and define the triangular array $Z_{Ti} \equiv \frac{1}{\sqrt{T}} G_T [I_T^{(\neg 0)}(\phi_T^*)]^{-1/2} \nabla^{1} \log \mathcal{L}_{Ti}(\phi_T^{*})$. Now, we verify Lindeberg's condition. Let $\varepsilon>0$ be an arbitrary constant, and note that by Cauchy-Schwarz
\begin{equation*}
\sum_{i=1}^{T} E\{\|Z_{Ti}\|^2 \textbf{1}_{(\|Z_{Ti}\|>\varepsilon)} \} \leq T E\{\|Z_{Ti}\|^4\}^{1/2} P\{\|Z_{Ti}\|>\varepsilon \}^{1/2}.
\end{equation*}
Then, by Chebyshev,
\begin{equation*}
P\{\|Z_{Ti}\|>\varepsilon \} \leq \frac{E\| G_T [I_T^{(\neg 0)}(\phi_T^*)]^{-1/2} \nabla^{1} \log \mathcal{L}_{Ti}(\phi_T^{*}) \|^2}{T \varepsilon^2}  = O(1/T),
\end{equation*}
where we have used the definition of the information matrix, and that $G_TG_T^\top \rightarrow G$.

Again by Cauchy-Schwarz, and the fact that the induced matrix norm of a symmetric matrix is equivalent to its spectral radius, we have
\begin{align*} 
E\{\|Z_{Ti}\|^4\} &\leq T^{-2} \lambda_{max}(G_TG_T^\top) \lambda_{max}(I_T^{(\neg 0)}(\phi_T^*)) E\| \nabla^{1} \log \mathcal{L}_{Ti}(\phi_T^{*}) [\nabla^{1} \log \mathcal{L}_{Ti}(\phi_T^{*})]^\top\|^2 \\
 &= O(K_T^{*2}/T^2),
\end{align*}
where $E\| \nabla^{1} \log \mathcal{L}_{Ti}(\phi_T^{*}) [\nabla^{1} \log \mathcal{L}_{Ti}(\phi_T^{*})]^\top\|^2$ has $K_T^{*2}$ bounded elements as implied by the proof of \thref{boundedmoments}. So we conclude that Lindeberg's condition is satisfied since $\sum_{i=1}^{T} E\{\|Z_{Ti}\|^2 \\ \textbf{1}_{(\|Z_{Ti}\|>\varepsilon)} \} = O(K_T^*/ \sqrt{T}) = o(1)$. Furthermore, for the asymptotic variance, we have $var(\sum_i^T Z_{Ti}) = T cov(Z_{Ti}) =  cov(G_T  [ I_T^{(\neg 0)} (\phi_T^*)]^{-1/2} \nabla^{1} \log \mathcal{L}_{Ti}(\phi_T^{*})) \rightarrow G$. Therefore, by the Lindeberg-Feller CLT, $\frac{1}{\sqrt{T}} G_T [I_T^{(\neg 0)}  (\phi_T^*)]^{-1/2}  \nabla^{1} \log  \mathcal{L}_{T}  (\phi_T^{*}) \rightarrow^{d} \mathcal{N}(0,G)$.

Again, under the assumptions \ref{itm:stat}-\ref{itm:minorization} we can use Lemma 2 of \cite{bickel1998asymptotic} to obtain $T^{-1}\nabla^{2} \log \mathcal{L}_T(\check{\phi}_T) \rightarrow^p - I_T^{\neg 0}(\phi_T^*)$ in \eqref{FOCunbiased}. Hence, putting all these results together in \eqref{FOCunbiased} completes the proof.

\qed

\end{document}


\thispagestyle{empty}

\begin{center}
{\LARGE Online appendix for "Estimating high-dimensional Markov-switching VARs"}\bigskip

{\large By Kenwin Maung}

\end{center}

\section{Proofs}

This section contains the proofs for the following results in the main text.

\subsection{Proof for \thref{momentresult}}

\begin{proof}
\eqref{infomat} follows from taking the derivative of \eqref{score}, so it suffices to show the latter. First, write $\nabla^{1} \log \mathcal{L}_T(\phi_T^*) = \sum^{T}_{t=p_T+1} \nabla^{1} \log \mathcal{L}(y_t| \mathcal{I}^{t-1}_{t-{p_T}}; \phi_T^*)$. By the Louis missing information principle \citep{louis1982finding}\footnote{See, for example, equation (5) in \cite{bickel1998asymptotic}}, we can express the first  derivative with respect to the $i^{th}$ element in $\phi_T^*$ as
\begin{align}\label{louis}
\nabla^{1}_i \log \mathcal{L}(y_t| \mathcal{I}^{t-1}_{t-{p_T}}; \phi_T^*) &= \sum_{k=t-p_T}^{t-1} \bigg\{E\bigg[\nabla^{1}_i \log p_{(S_{k-1}) \shortrightarrow (S_{k})}(\phi_T^*) + \nabla^{1}_i \log g(y_k| \mathcal{I}^{k-1}_{k-p_T}; \Phi_{S_k,T}^*) \bigg| \mathcal{I}^{t}_{t-p_T}\bigg] \notag \\ 
&- E\bigg[ \nabla^{1}_i \log p_{(S_{k-1}) \shortrightarrow (S_{k})}(\phi_T^*) + \nabla^{1}_i \log g(y_k| \mathcal{I}^{k-1}_{k-p_T}; \Phi_{S_k,T}^*)\bigg| \mathcal{I}^{t-1}_{t-p_T}\bigg] \bigg\} \notag \\
&+ E\bigg[ \nabla^{1}_i \log g(y_t| \mathcal{I}^{t-1}_{t-p_T}; \Phi_{S_t,T}^*) \bigg| \mathcal{I}^{t}_{t-p_T} \bigg] \notag \\
&+ E\bigg[\nabla^{1}_i \log p_{t-p_T}(\phi_T^*) \bigg| \mathcal{I}^{t}_{t-p_T} \bigg] - E\bigg[\nabla^{1}_i \log p_{t-p_T}(\phi_T^*) \bigg| \mathcal{I}^{t-1}_{t-p_T} \bigg].
\end{align}
Then, using \eqref{louis}, the law of iterated expectation, and the assumption that $g(y_t|\cdot)$ is a valid probability density function, we have that $E[\nabla^{1}_i \log \mathcal{L}(y_t|  \mathcal{I}^{t-1}_{t-{p_T}}; \phi_T^*) | \mathcal{I}^{t-1}_{t-{p_T}}] = 0$, and therefore \eqref{score} follows.  
\end{proof}

\subsection{Proof for \thref{boundedmoments}}

\begin{proof}
Under assumptions \ref{itm:stat}-\ref{itm:minorization}, we can employ Lemmas 3-6 from \cite{bickel1998asymptotic}, with some modification to the conditional density, to show that $\nabla^{1}_j \mathcal{L}_T(\phi_T^*) \rightarrow 0$ in $L_2$. Likewise, Lemmas 7-9 of \cite{bickel1998asymptotic} also imply that $\nabla^{2}_{j,k} \mathcal{L}_T(\phi_T^*)$ is bounded in $L_1$. However, from the proofs of Lemmas 7-9, we can see that this result will extend to $L_2$.

Hence, here we show the result for the third derivative. Similar to what we did in \thref{momentresult}, write $\nabla^{3} \log \mathcal{L}_T(\phi_T) = \sum_{t=p_T+1}^T \nabla^{3} \log \mathcal{L}(y_t| \mathcal{I}_{t-p_T}^{t-1}; \phi_T)$. Again, by the Louis missing information principle, we can get
\begin{align}
&\nabla^{3}_{j,k,l} \log \mathcal{L}(y_t| \mathcal{I}_{t-p_T}^{t-1}; \phi_T) \notag \\ 
&= \nabla^{3}_{j,k,l} \log \mathcal{L}(\underbrace{y_{t-p_T},\ldots, y_{t}}_{\equiv y_{t-p_T}^{t}}| \mathcal{X}_{t-p_T}^{t-1}; \phi_T) - \nabla^{3}_{j,k,l} \log \mathcal{L}(\underbrace{y_{t-p_T},\ldots, y_{t-1}}_{\equiv y_{t-p_T}^{t-1}}| \mathcal{X}_{t-p_T}^{t-1}; \phi_T), \label{louisthird1}
\end{align}
where $\mathcal{X}_{t-p_T}^{t-1} = (x_{t-q_T}, \ldots, x_{t-1})$,
\begin{align}\label{louisthird2}
\nabla^{3}_{j,k,l} &\log \mathcal{L}(y_{t-p_T}^{t}| \mathcal{X}_{t-p_T}^{t-1}; \phi_T) \notag \\ &= E[\nabla^{3}_{j,k,l} \log \mathcal{L}(y_{t-p_T}^{t}, S_{t-p_T}^{t}| \mathcal{X}_{t-p_T}^{t-1}; \phi_T) | \mathcal{I}_{t-p_T}^{t}] \notag \\
&+ 2 E[\nabla^{2}_{j,k} \log \mathcal{L}(y_{t-p_T}^{t}, S_{t-p_T}^{t}| \mathcal{X}_{t-p_T}^{t-1}; \phi_T) |  \mathcal{I}_{t-p_T}^{t}] \notag \\
&- 2 E[\nabla^{1}_{j} \log \mathcal{L}(y_{t-p_T}^{t}, S_{t-p_T}^{t}| \mathcal{X}_{t-p_T}^{t-1}; \phi_T) |  \mathcal{I}_{t-p_T}^{t}] \times E[\nabla^{2}_{j,k} \log \mathcal{L}(y_{t-p_T}^{t}, S_{t-p_T}^{t}| \mathcal{X}_{t-p_T}^{t-1}; \phi_T) |  \mathcal{I}_{t-p_T}^{t}], 
\end{align}
and $S_{a}^{b} = (S_a \ldots, S_b)$. Note that the last two terms are bounded in $L_2$ because $\nabla_{j,k}^2 \mathcal{L}_T(\phi_T)$ can be expressed as a function of the conditional expectations of first and second derivatives similar to \eqref{louisthird2} \citep[see (14) in][]{kasahara2019asymptotic}, and it is bounded in $L_2$. So we only have to show for the first term (label this $E[\nabla^3|\mathcal{I}_{t-p_T}^{t}]$). Fortunately, this term results in a similar expression to \eqref{louis}. From \eqref{louisthird1},
\begin{align}
&E[\nabla^3|\mathcal{I}_{t-p_T}^{t}] - E[\nabla^3|\mathcal{I}_{t-p_T}^{t-1}] \notag \\ 
&= \sum_{n=t-p_T}^{t-1} \bigg\{E\bigg[\nabla^{3}_{j,k,l} \log p_{(S_{n-1}) \shortrightarrow (S_{n})}(\phi_T) + \nabla^{3}_{j,k,l} \log g(y_n| \mathcal{I}^{n-1}_{n-p_T}; \Phi_{S_n,T}) \bigg| \mathcal{I}^{t}_{t-p_T}\bigg] \notag \\ 
&- E\bigg[ \nabla^{3}_{j,k,l} \log p_{(S_{n-1}) \shortrightarrow (S_{n})}(\phi_T) + \nabla^{3}_{j,k,l} \log g(y_n| \mathcal{I}^{n-1}_{n-p_T}; \Phi_{S_n,T})\bigg| \mathcal{I}^{t-1}_{t-p_T}\bigg] \bigg\} \notag \\
&+ E\bigg[ \nabla^{3}_{j,k,l} \log g(y_t| \mathcal{I}^{t-1}_{t-p_T}; \Phi_{S_t,T}) \bigg| \mathcal{I}^{t}_{t-p_T} \bigg] \notag \\
&+ E\bigg[\nabla^{3}_{j,k,l} \log p_{t-p_T}(\phi_T) \bigg| \mathcal{I}^{t}_{t-p_T} \bigg] - E\bigg[\nabla^{3}_{j,k,l} \log p_{t-p_T}(\phi_T) \bigg| \mathcal{I}^{t-1}_{t-p_T} \bigg].
\end{align}
With this expression, the proof is completed if we can show the following:

\begin{enumerate}[label=(\roman*)]
\item $\big\| E[\nabla^{3}_{j,k,l} \log g(y_n| \mathcal{I}^{n-1}_{n-p_T}; \Phi_{S_n,T})| \mathcal{I}^{t}_{t-p_T}] - E[\nabla^{3}_{j,k,l} \log g(y_n| \mathcal{I}^{n-1}_{n-p_T}; \Phi_{S_n,T})| \mathcal{I}^{t-1}_{t-p_T}] \big\|_2 \leq K \beta^{T}$,

\item $\big\| E[\nabla^{3}_{j,k,l} \log p_{(S_{n-1}) \shortrightarrow (S_{n})}(\phi_T) | \mathcal{I}^{t}_{t-p_T}] - E[\nabla^{3}_{j,k,l} \log p_{(S_{n-1}) \shortrightarrow (S_{n})}(\phi_T) | \mathcal{I}^{t-1}_{t-p_T}] \big\|_2 \leq K \beta^{T}$,

\item $\big\| E[\nabla^{3}_{j,k,l} \log p_{t-p_T}(\phi_T) | \mathcal{I}^{t}_{t-p_T}] - E[\nabla^{3}_{j,k,l} \log p_{t-p_T}(\phi_T) | \mathcal{I}^{t-1}_{t-p_T}] \big\|_2 \leq K \beta^{T}$,
\end{enumerate}
where $K >0$ is constant, and $\beta \in [0,1)$.

This can be shown with a similar approach to that of lemma 6 in \cite{bickel1998asymptotic}. Under \ref{itm:stat} and \ref{itm:cont}, there exists a constant $\nu > 0$ such that $\inf\{p_{(s)\shortrightarrow(s^{'})}(\phi_T)|s,s^{'},\phi_T \in \Theta_T\} \geq \nu$ and $\inf\{p_{s}(\phi_T)|s,\phi_T \in \Theta_T\} \geq \nu$. Define $\mu(y_t) = [1 + (M-1)\nu^{-2}\rho(y_t)]^{-1}$ where $\rho(\cdot)$ is defined in \ref{itm:minorization}. We show the result for (i) only, while the rest follow similarly.
\begin{align*}
&| E[\nabla^{3}_{j,k,l} \log g(y_n| \mathcal{I}^{n-1}_{n-p_T}; \Phi_{S_n,T})| \mathcal{I}^{t}_{t-p_T}] - E[\nabla^{3}_{j,k,l} \log g(y_n| \mathcal{I}^{n-1}_{n-p_T}; \Phi_{S_n,T})| \mathcal{I}^{t-1}_{t-p_T}]| \\
&= \bigg|\sum_{s=1}^M \nabla^{3}_{j,k,l} \log g(y_n| \mathcal{I}^{n-1}_{n-p_T}; \Phi_{s,T}) \{P(S_n = s| \mathcal{I}^{t}_{t-p_T}) - P(S_n = s| \mathcal{I}^{t-1}_{t-p_T})\}\bigg| \\
&\leq \max_s |\nabla^{3}_{j,k,l} \log g(y_n| \mathcal{I}^{n-1}_{n-p_T}; \Phi_{s,T})| K \prod_{i=n+1}^{t-1} \exp(-2\mu(y_i)),
\end{align*}
for some constant $K$, and the last inequality follows from lemma 5 in \cite{bickel1998asymptotic}. Then, observe that
\begin{align*}
&\| E[\nabla^{3}_{j,k,l} \log g(y_n| \mathcal{I}^{n-1}_{n-p_T}; \Phi_{S_n,T})| \mathcal{I}^{t}_{t-p_T}] - E[\nabla^{3}_{j,k,l} \log g(y_n| \mathcal{I}^{n-1}_{n-p_T}; \Phi_{S_n,T})| \mathcal{I}^{t-1}_{t-p_T}]\|_2^2 \\
&\leq K E\bigg[ \max_s |\nabla^{3}_{j,k,l} \log g(y_n| \mathcal{I}^{n-1}_{n-p_T}; \Phi_{s,T})|^2 \prod_{i=n+1}^{t-1} \exp(-4\mu(y_i)) \bigg] \\
&= K E\bigg[ E\bigg[\max_s |\nabla^{3}_{j,k,l} \log g(y_n| \mathcal{I}^{n-1}_{n-p_T}; \Phi_{s,T})|^2 \prod_{i=n+1}^{t-1} \exp(-4\mu(y_i)) \bigg| S_{t-p_T+1}^{n} \bigg]\bigg] \\
&= K E \bigg[ E[\max_s |\nabla^{3}_{j,k,l} \log g(y_n| \mathcal{I}^{n-1}_{n-p_T}; \Phi_{s,T})|^2 | S_n] \prod_{i=n+1}^{t-1} E[\exp(-4\mu(y_i)) |S_i]\bigg] \\
&\leq K \max_{s^{'}} E[\max_s |\nabla^{3}_{j,k,l} \log g(y_n| \mathcal{I}^{n-1}_{n-p_T}; \Phi_{s,T})|^2 | S_n = s^{'}] \beta^{t-n-1}
\end{align*}
where we have used the fact that $0<\mu(\cdot)<1$ and the property of the exponential to bind it with $\beta \in [0,1)$. Together with \ref{itm:momentcondition}, we get the result in (i). 
\end{proof}

\subsection{Proof for \thref{Oprate}}

\begin{proof} The proof is similar to the proof of Lemma A.1 in \cite{kwon2012large}. 

For notational convenience, let $(A)_{ij}$ indicate the element in the $i^{th}$ row and $j^{th}$ column of $A$, $\nabla^{1} \log \mathcal{L}_T(\phi_T^*) \equiv \nabla^{1} \mathcal{L}$, and $\nabla^{2}\mathcal{L}_T(\phi_T^*) - E[\nabla^{2}\mathcal{L}_T(\phi_T^*)] \equiv \nabla^{2} \mathcal{L}$.  

For (1),
\begin{equation*} 
E[ \| \nabla^{1} \mathcal{L} \| ^ 2]  =  \sum_{i=1}^{K_T^*} E[(\nabla^{1} \mathcal{L})_{i}^2] = O(T K_T^*),
\end{equation*}
where the last equality is an application of Cauchy-Schwarz, and the $L_2$ bound in \thref{boundedmoments}. Then, (2) follows from Chebyshev's inequality.

The proof of (2) is almost identical except for the fact that we are dealing with a matrix so that
\begin{equation*}
E[\|\nabla^2 \mathcal{L} \|_{\textbf{1}}^2] = \sum_{i=1}^{K_T^*} \sum_{j=1}^{K_T^*} E[(\nabla^2\mathcal{L})_{ij}^2] = O(T K_T^{*2}).
\end{equation*}

(3) follows similarly from (2) since it it just the coordinate-wise version. Likewise, the proof for (4) is identical but with a triple sum instead, and (5) follows.
\end{proof} 

\newpage

\bibliographystyle{elsarticle-harv}
\bibliography{third_year_paper}

\appendix